\documentclass[11pt,a4paper]{article}
\usepackage[margin=1in]{geometry}
\usepackage{amsmath,amssymb,amsthm,color,enumerate,graphicx,stackrel,natbib,authblk,setspace}
\usepackage{lscape}
\usepackage{placeins}

\usepackage{epstopdf}
\usepackage{subfigure}
\usepackage{multirow}

\begin{document}


\title{Pricing methods for $\alpha$-quantile and perpetual early exercise options based on Spitzer identities}

\author{Carolyn E. Phelan

Financial Computing and Analytics Group, Department of Computer Science, University College London

\texttt{c.phelan@cs.ucl.ac.uk}

\bigskip

Daniele Marazzina

Department of Mathematics, Politecnico di Milano

\texttt{daniele.marazzina@polimi.it}

\bigskip

Guido Germano

Financial Computing and Analytics Group, Department of Computer Science, University College London\\ Systemic Risk Centre, London School of Economics

\texttt{g.germano@ucl.ac.uk, g.germano@lse.ac.uk}
}

\maketitle

\begin{abstract}
We present new numerical schemes for pricing perpetual Bermudan and American options as well as $\alpha$-quantile options. This includes a new direct calculation of the optimal exercise barrier for early-exercise options. Our approach is based on the Spitzer identities for general L\'evy processes and on the Wiener-Hopf method. Our direct calculation of the price of $\alpha$-quantile options combines for the first time the Dassios-Port-Wendel identity and the Spitzer identities for the extrema of processes. Our results show that the new pricing methods provide excellent error convergence with respect to computational time when implemented with a range of L\'evy processes.
\end{abstract}


\section{Introduction}
Much of the recent literature in finance concerns exotic options where the payoff depends on the path of the underlying asset.
Fluctuation identities such as the ones by Spitzer 
have applications in the pricing of many of these contracts. 
Here, we introduce novel methods for pricing perpetual Bermudan and American options, and $\alpha$-quantile options, using these identities.

Option pricing is a classic problem in financial management and has been subject to many different approaches such as the one by \cite{Tse2001} for discretely monitored hindsight and barrier options. However, this approach was limited to Brownian motion and, as demonstrated by \cite{kou2002jump} and \cite{Du2017}, a process with both jump and diffusion processes is a more appropriate model. The method we outline here is based on probabilistic identities and, as it is valid for general L\'evy processes, it is applicable to a wide range of jump-diffusion models. Furthermore, many decision-making processes can be modelled mathematically as American options \citep{Battauz2015} and thus the new methods we present here also have a more general application in finance.

While European options can only be exercised at a single maturity, American options can be exercised at any time up to expiry. Bermudan options are halfway between the two in that they can be exercised at a finite set of dates, i.e they are a discretely monitored version of American options. These can have an expiry date, beyond which the contract is worthless, or no expiry date, in which case they are called perpetual options. The valuation of American options is a long-standing problem in mathematical finance \citep{Merton1973, Brennan1977} as it combines a pricing problem with an optimisation problem (i.e the computation of the early exercise barrier) and a pricing problem. A closed-form solution has not been found for a finite expiry date.

In contrast, perpetual American options permit a closed-form solution when the underlying is driven by geometric Brownian motion \citep{Merton1973}, as the perpetual nature of the option means that the optimal exercise barrier is constant rather than a function of time to expiry. However, this closed-form solution cannot be extended to Bermudan options. Moreover, as explained by \cite{Boyarchenko2002}, the smooth pasting method used to price perpetual American options can fail under jump processes such as those in the general L\'evy class.

Several approximate methods suggested for fixed-expiry American options, such as finite differences \citep{Brennan1977}, trees \citep{Cox1979}, Monte Carlo \citep{Rogers2002} and recursive Hilbert transforms \citep{Feng2013} are inherently discrete and thus lend themselves to Bermudan options with finite expiry. However, they are not particularly accurate or efficient for perpetual Bermudan options as the computational load increases with the number of monitoring dates which is, of course, infinite for these types of contract. Moreover, in times of very low interest rates such as the last decade, the existence of exercise dates even many years in the future can have a significant effect on the option value and thus the number of dates cannot generally be truncated to a sufficiently low value for these methods to be computationally efficient for perpetual options.


\cite{Boyarchenko2002} published a method to price perpetual American options for many L\'evy processes using analytic approximations of the Wiener-Hopf factors. This was a step forward in showing the applicability of the Wiener-Hopf method to price perpetual options with L\'evy processes. However, it is an approximate solution as there is no general closed-form method to calculate the Wiener-Hopf factors; in addition, the proposed method is not applicable to all L\'evy processes and specifically excludes the variance gamma (VG) process. In \cite{Boyarchenko2002_2} this method was adapted to perpetual Bermudan options; however, the Wiener-Hopf factorisation again requires approximation in some cases and calculations were presented for simple jump-diffusion and normal inverse Gaussian (NIG) processes only. 

\cite{Mordecki2002} also devised a pricing approximation for perpetual American put options with L\'evy processes which is based on the optimal stopping problem for partial sums by \cite{Darling1972}; therefore it intrinsically operates in discrete time and thus is useful for Bermudan options. However, this method has restrictions on the jump measure used in the L\'evy-Khinchine representation of the characteristic function and therefore cannot be used for general L\'evy processes. 

Here we fill the gap in the literature for a direct numerical pricing method for perpetual Bermudan and American options which can be used for general L\'evy models. In Section \ref{sec:7_pbo} we describe a novel a pricing scheme based on the Spitzer identities which includes a new way to directly calculate the optimal exercise barrier. We also provide the first implementation of the method by \cite{GreenThesis2009} based on the Spitzer identities and an expression for prices of first-touch and overshoot options which uses residue calculus. 

In this article we also deal with quantile options which belong to the class of hindsight options. These have a fixed expiry where the payoff at expiry is determined by the path of the option up to that date. Two such examples are lookback options, as priced by \cite{fusai2015}, and quantile options. Fixed-strike lookback options have a similar payoff to European options except that, instead of being a function of the underlying asset price at expiry, it uses the maximum or minimum over the monitoring period. Quantile options can be considered an extension of lookback options because, rather than using the maximum or minimum of an asset price, the payoff is based on the value which the asset price spends $\alpha$\% of the time above or below. For this reason they are often described as $\alpha$-quantile options. 

They were first suggested by \cite{Miura1992} as a way to design a hindsight option which is less sensitive to very extreme, but short lived, phenomena in the asset price process compared to simple lookback options. \cite{Akahori1995} and \cite{Yor1995} published analytic methods for pricing these contracts with Brownian motion. Since then, most work on pricing these types of options has been based on the remarkable identity by \cite{dassios1995} which used work by \cite{Wendel1960} and \cite{Port1963} and relates the probability distribution of the $\alpha$-quantile to the probability distribution of the maximum and minimum of two independent processes. A note by \cite{dassios2006quantiles} showed that this can also be extended to general L\'evy processes. Several solutions for L\'evy processes have been developed, such as Monte-Carlo methods for jump-diffusion processes by \cite{Ballotta2002} and an analytic method for the \cite{kou2002jump} double-exponential process by \cite{Cai2010}. For discrete monitoring, \cite{fusai2001pricing} explored the relationship between continuously and discretely monitored $\alpha$-quantile options however, whilst they made recommendations for selecting an optimal pricing method depending on the value of the monitoring interval $\Delta t$, they did not produce a correction term like the one by \cite{broadie1997} for barrier options.

\cite{atkinson2007discrete} provided closed-form prices for discretely monitored quantile options by using the $z$-transform to write the problem in terms of a Wiener-Hopf equation. They solved this analytically for prices driven by geometric Brownian motion and also demonstrated the relationship between their results and the Spitzer identities. This approach was extended to general L\'evy processes by \cite{GreenThesis2009}, who developed direct methods, based on his formulation of the Spitzer identities, for calculating the distribution of the supremum and infimum of processes. These methods were implemented for lookback options in \cite{fusai2015} which showed exponential convergence for general L\'evy processes with a CPU time which is independent of the number of monitoring dates. In this paper we address the gap in the literature for a general method for pricing $\alpha$-quantile options with general L\'evy processes and a date-independent computational time. 

The article is organised as follows: Section \ref{sec:fourhilb} provides the background to transform techniques, with special reference to the use of the Fourier transform in option pricing. Sections \ref{sec:Quantile}, \ref{sec:7_pbo} and \ref{sec:PerAm} describe the new techniques for pricing $\alpha$-quantile, perpetual Bermudan and perpetual American options and present results for these new methods for a range of L\'evy processes. All the numerical results were obtained using MATLAB R2016b running under OS X Yosemite on a 2015 Retina MacBook Pro with a 2.7GHz Intel Core i5 processor and 8GB of RAM following the step-by-step procedures described in Appendix \ref{sec:App_proc} and using the process parameters listed in Appendix \ref{sec:App_params}.

\section{Transform methods for option pricing}\label{sec:fourhilb}

In this paper we make extensive use of the Fourier transform \citep[see e.g.][]{Polyanin1998,Kreyszig2011}, an integral transform with many applications. Historically, it has been widely employed in spectroscopy and communications, therefore much of the literature refers to the function in the Fourier domain as its spectrum. According to the usual convention in financial literature, we define the forward and inverse Fourier transforms as
\begin{align}
\widehat{f}(\xi)&=\mathcal{F}_{x\rightarrow\xi} \big[f(x)\big]=\int^{+\infty}_{-\infty}e^{i\xi x}f(x)dx, \label{eq:FwdFourier}\\
f(x)&=\mathcal{F}^{-1}_{\xi\rightarrow x} \big[\widehat{f}(\xi)\big]=\frac{1}{2\pi}\int^{+\infty}_{-\infty}e^{-i\xi x}\widehat{f}(\xi)d\xi \label{eq:RevFourier}.
\end{align}

Let $S(t)$ be the price of an underlying asset and $x(t) = \log(S(t)/S_0)$ its (rescaled) log-price. To find the price $v(x,t)$ of an option at time $t=0$ when the initial price of the underlying is $S(0) = S_0$, and thus its log-price is $x(0)=0$, we need to discount the expected value of the undamped option payoff $\phi(x(T))e^{-\alpha_{\mathrm{d}} x(T)}$ at maturity $t=T$ with respect to an appropriate risk-neutral probability distribution function (PDF) $p(x,T)$ whose initial condition is $p(x,0) = \delta(x)$. Here $\alpha_{\mathrm{d}}$ is the damping parameter, the selection of which was described in more details by \cite{Feng2008}. As shown by
\cite{lewis2001simple}, and comprehensively used in \cite{fusai2015} and \cite{Phelan2017}, this can be done using the Plancherel relation,
\begin{align}
v(0,0) & = e^{-rT}\mathrm{E}\left[\phi(x(T))e^{-\alpha_{\mathrm{d}} x(T)}|x(0)=0\right]=e^{-rT}\int^{+\infty}_{-\infty}\phi(x)e^{-\alpha_{\mathrm{d}} x}p(x,T)dx \nonumber\\
& = \frac{e^{-rT}}{2\pi}\int^{+\infty}_{-\infty}\widehat{\phi}(\xi)\widehat{p}^{\,*}(\xi+i\alpha_{\mathrm{d}},T)d\xi = e^{-rT}\mathcal{F}^{-1}_{\xi\rightarrow x}\left[\widehat{\phi}(\xi)\widehat{p}^{\,*}(\xi+i\alpha_{\mathrm{d}},T)\right](0). \label{eq:Planch}
\end{align}
Here, $\widehat{p}^{\,*}(\xi+i\alpha_{\mathrm{d}},T)$ is the complex conjugate of the Fourier transform of $e^{-\alpha_{\mathrm{d}} x}p(x,T)$. To price options using this relation, we need the Fourier transforms of both the damped payoff and the PDF.
A double-barrier option has the damped payoff
\begin{equation}
\label{eq:damped_payoff}
\phi(x) = e^{\alpha_{\mathrm{d}} x}S_0(\theta(e^x-e^k))^+\mathbf{1}_{(l,u)}(x),
\end{equation}
where $e^{\alpha_{\mathrm{d}} x}$ is the damping factor, $\theta = 1$ for a call, $\theta = -1$ for a put, $\mathbf{1}_A(x)$ is the indicator function of the set $A$, $k=\log(K/S_0)$ is the log-strike, $u=\log(U/S_0)$ is the upper log-barrier, $l=\log(L/S_0)$ is the lower log-barrier, $K$ is the strike price, $U$ is the upper barrier and $L$ is the lower barrier. The Fourier transform of the damped payoff $\phi(x)$ is available analytically, 
\begin{equation}
\label{eq:Payoff}
\widehat{\phi}(\xi)=S_0\left(\frac{e^{(1+i\xi+\alpha_{\mathrm{d}})a}-e^{(1+i\xi+\alpha_{\mathrm{d}})b}}{1+i\xi+\alpha_{\mathrm{d}}}-\frac{e^{k+(i\xi+\alpha_{\mathrm{d}})a}-e^{k+(i\xi+\alpha_{\mathrm{d}})b}}{i\xi+\alpha_{\mathrm{d}}}\right),
\end{equation}
where for a call option $a = u$ and $b = \max(k,l)$, while for a put option $a=l$ and $b=\min(k,u)$.

The Fourier transform of the PDF $p(x,t)$ of a stochastic process $X(t)$ is the characteristic function
\begin{equation}
\label{eq:CharFun}
\Psi(\xi,t)=\mathrm{E}\left[e^{i\xi X(t)}\right]=\int^{+\infty}_{-\infty}e^{i\xi x}p(x,t)dx=\mathcal{F}_{x\rightarrow\xi}\left[p(x,t)\right]=\widehat{p}(\xi,t).
\end{equation}
For a L\'evy process the characteristic function can be written as $\Psi(\xi,t)=e^{\psi(\xi)t}$, where $\psi(\xi)$ is known as the characteristic exponent and uniquely defines the process. 

The fluctuation identities we use in the pricing methods for discretely monitored options are expressed in the Fourier-$z$, with the Fourier transform applied to the log-price domain and the $z$-transform applied to the discrete monitoring times. The $z$-transform is defined as
\begin{align}
\mathcal{Z}_{n\rightarrow q}[f(n)]:=\sum_{n=0}^\infty q^n f(n), \quad q\in\mathbb{C},
\end{align}
where $f(n)=f(t_n),\ n\in\mathbb{N}_0$, is a function of a discrete variable. For some of the pricing methods we also require the inverse $z$-transform. This is not generally available in closed form, so we use the well established method by \cite{Abate1992} which has been used for option pricing by \cite{fusai2006exact,fusai2015,Phelan2017}.

For continuously monitored options the Laplace transform is applied to the time domain. 
The forward Laplace transform is an integral transform similar to the Fourier transform and it is
\begin{align}
	\tilde{f}(s)=\mathcal{L}_{t\rightarrow s}\big[f(t)\big]=\int^{\infty}_{0}f(t)e^{-st}dt. \label{eq:Laplace}
\end{align}
Similarly to the discrete time case, the inverse transform is not generally available in closed form and so we use the numerical inverse Laplace transform by \cite{Abate1995}, used for option pricing by \cite{Phelan2017Fluct}.
The relationship between the forward Laplace and $z$ transforms is
\begin{align}
\label{eq:2_an_ztolap}
\mathcal{L}_{t\rightarrow s}[f(t)]&=\lim_{\Delta t\rightarrow0}\Delta t \mathcal{Z}\left[f(t_n)\right].
\end{align}
This relationship can be exploited to derive versions of the fluctuation identities with continuous monitoring \citep{baxter1957distribution,Green2010,fusai2015,Phelan2017Fluct}.
\subsection{Decomposition and factorisation}\label{sec:decomfact}
The fluctuation identities require the decomposition and factorisation of functions in the Fourier domain. Decomposition splits a function in the Fourier domain into two parts which are non-zero in the log-price domain only above or below a fixed value respectively, i.e.
\begin{align}
\widehat{f}(\xi)=\widehat{f_{l+}}(\xi)+\widehat{f_{l-}}(\xi),
\end{align}
where $\widehat{f_{l+}}(\xi)=\mathcal{F}_{x\rightarrow\xi}[\mathbf{1}_{(l,\infty)}(x)\mathcal{F}^{-1}_{\xi\rightarrow x}[\widehat{f}(\xi)]]$ and $\widehat{f_{l-}}(\xi)=\mathcal{F}_{x\rightarrow\xi}[\mathbf{1}_{(-\infty,l)}(x)\mathcal{F}^{-1}_{\xi\rightarrow x}[\widehat{f}(\xi)]]$. 
We use the Plemelj-Sokhotsky relations for the decomposition,
\begin{align}
& \widehat{f_{l+}}(\xi)=\frac{1}{2}\big\{\widehat{f}(\xi)+e^{il\xi}i\mathcal{H}\big[e^{-il\xi}\widehat{f}(\xi)\big]\big\}\label{eq:2_an_PSRelgenpos}\\
& \widehat{f_{l-}}(\xi)=\frac{1}{2}\big\{\widehat{f}(\xi)-e^{il\xi}i\mathcal{H}\big[e^{-il\xi}\widehat{f}(\xi)\big]\big\}.\label{eq:2_an_PSRelgenneg}
\end{align}
where $\mathcal{H}[\cdot]$ denotes the Hilbert transform, which is implemented numerically using the sinc method described by \cite{Stenger1993} and used by \cite{Feng2008} and \cite{fusai2015} for option pricing. The latter paper describes the implementation of this method to compute fluctuation identities for option pricing; a detailed error analysis of the technique within this application was subsequently carried out by \cite{Phelan2017Fluct,Phelan2017}. The calculations also require the factorisation of a function, i.e obtaining $\widehat{g_\oplus}(\xi)$ and $\widehat{g_\ominus}(\xi)$ such that $\widehat{g}(\xi)=\widehat{g_\oplus}(\xi)\widehat{g_\ominus}(\xi)$. This is done by decomposing the logarithm of $\widehat{g}(\xi)$ and then exponentiating the result.
\section{Quantile options}\label{sec:Quantile}
The $\alpha$-quantile $X_\alpha$ of a random process $X(t)$ is the value which it is below $\alpha$\% of the time; $\alpha$-quantile options have a payoff which depends on this value. They were designed by \cite{Miura1992} as a variation on lookback options which would be less susceptible to the effects of very large, but short lived, swings in the price of the underlying asset. They are therefore more resistant to market manipulation as it is easier to cause a brief swing in an asset price compared to a longer term price movement. We propose a pricing method for $\alpha$-quantile options with general exponential L\'evy processes.
\subsection{Pricing discretely monitored quantile options}
For $\alpha$-quantile options the form of the payoff is the same as in Eq.~(\ref{eq:damped_payoff}), 
but in this case it is calculated as a function of the quantile, i.e. $x=X_\alpha$ rather than the process value at expiry $x=X(T)$. Therefore, with the characteristic function of $X_\alpha$, we can price an $\alpha$-quantile option in the Fourier domain using the Plancherel relation in Eq.~(\ref{eq:Planch}) and the Fourier transform of the damped payoff $\widehat{\phi}(\xi)$ from Eq.~(\ref{eq:Payoff}).

The Dassios-Port-Wendel identity \citep{Wendel1960,Port1963,dassios1995} states that the $\alpha$-quantile of a Brownian motion over the interval $[0,T]$ has the same distribution as the sum of the infimum of a Brownian motion over the time $[0,1-\alpha T]$ and the supremum of an independent Brownian motion over the time $[0,\alpha T]$. That is if $X_m=\min_{t\in[0,(1-\alpha) T]}X_{\mathrm{1}}(t)$ and $X_M=\max_{t\in[0,\alpha T]}X_{\mathrm{2}}(t)$, where $X_{\mathrm{1}}(t)$ and $X_{\mathrm{2}}(t)$ are independent Brownian motions, then
\begin{equation}
X_\alpha\stackrel{d}{=}X_m+X_M\label{eq:7_qu_DPW},
\end{equation}
where $X_\alpha$ is the $\alpha$-quantile of the Brownian motion.

We can understand the link between the value of $X_\alpha$ and the supremum and infimum intuitively. Firstly, if a process is split into a section for $t\in[0,\alpha T]$ and one for $t\in(\alpha T,T]$ then, by the property of independent increments, they represent two independent processes over $t\in[0,\alpha T]$ and $t\in(0,(1-\alpha)T]$. Moreover, the supremum of $X_\mathrm{1}(t)$ is the value that the process has spent $\alpha T$ time below and conversely the infimum of $X_\mathrm{2}(t)$ is the value which this process has spent $(1-\alpha)T$ time above. Although the basic idea behind the relationship in Eq.~(\ref{eq:7_qu_DPW}) is quite clear, the mathematical proof is quite involved and we refer the interested reader to the original paper by \cite{dassios1995} for the details. A note by \cite{dassios2006quantiles} showed that this identity also extends to general L\'evy processes.

\cite{Green2010} devised Spitzer-based formulations for the probability distributions of the maximum and minimum of a process which were used by \cite{fusai2015} to price fixed-strike lookback options with exponential error convergence. These are defined in the Fourier-$z$ domain as
\begin{align}
\widetilde{\widehat{p}}_{X_M}(\xi,q)=\frac{1}{\Phi_\oplus(\xi,q)\Phi_\ominus(0,q)} \label{eq:7_qu_charxmax}\\
\widetilde{\widehat{p}}_{X_m}(\xi,q)=\frac{1}{\Phi_\oplus(0,q)\Phi_\ominus(\xi,q)}, \label{eq:7_qu_charxmin}
\end{align}
where $\Phi_\oplus(\xi,q)$ and $\Phi_\ominus(\xi,q)$ are the Wiener-Hopf factors of $1-q\Psi(\xi,\Delta t)$ as described in Section \ref{sec:decomfact}. The inverse $z$-transform can be applied to obtain 
\begin{gather}
\widehat{p}_{X_M}(\xi,j)=\mathcal{Z}^{-1}_{q\rightarrow j} [\widetilde{\widehat{p}}_{X_M}(\xi,q)],\label{eq:7_qu_charxmaxinvz}\\
\widehat{p}_{X_m}(\xi,N-j)=\mathcal{Z}^{-1}_{q\rightarrow N-j}[\widetilde{\widehat{p}}_{X_m}(\xi,q)],\label{eq:7_qu_charxmininvz}
\end{gather}
in the Fourier-domain, where $N$ is the number of discrete monitoring dates and $j$ is the approximation of $\alpha N$ to the nearest integer, although contracts are often written to avoid this approximation by selecting $\alpha$ and $N$ so that $\alpha N$ is an integer. As $X_M$ and $X_m$ are the maximum and minimum of mutually independent processes, they are mutually independent random variables. It is a basic result in probability theory that the PDF of the sum of two independent random variables is equal to the convolution of their PDFs, i.e.\ 
\begin{align}
p_{X_\alpha}(x)=\int^{+\infty}_{-\infty}p_{X_M}(x')p_{X_m}(x-x')dx';
\end{align}
by convolution theorem this becomes a plain product in the Fourier domain,
\begin{align}
\widehat{p}_{X_\alpha}(\xi)=\widehat{p}_{X_M}(\xi,j)\widehat{p}_{X_m}(\xi,N-j).
\end{align}
The option price can then be obtained from the Plancherel relation in Eq.~(\ref{eq:Planch}) using the Fourier transform of the damped payoff function $\widehat{\phi}(\xi)$ from Eq.~(\ref{eq:Payoff}).

\begin{figure}
\begin{center}
\includegraphics[width=0.48\textwidth]{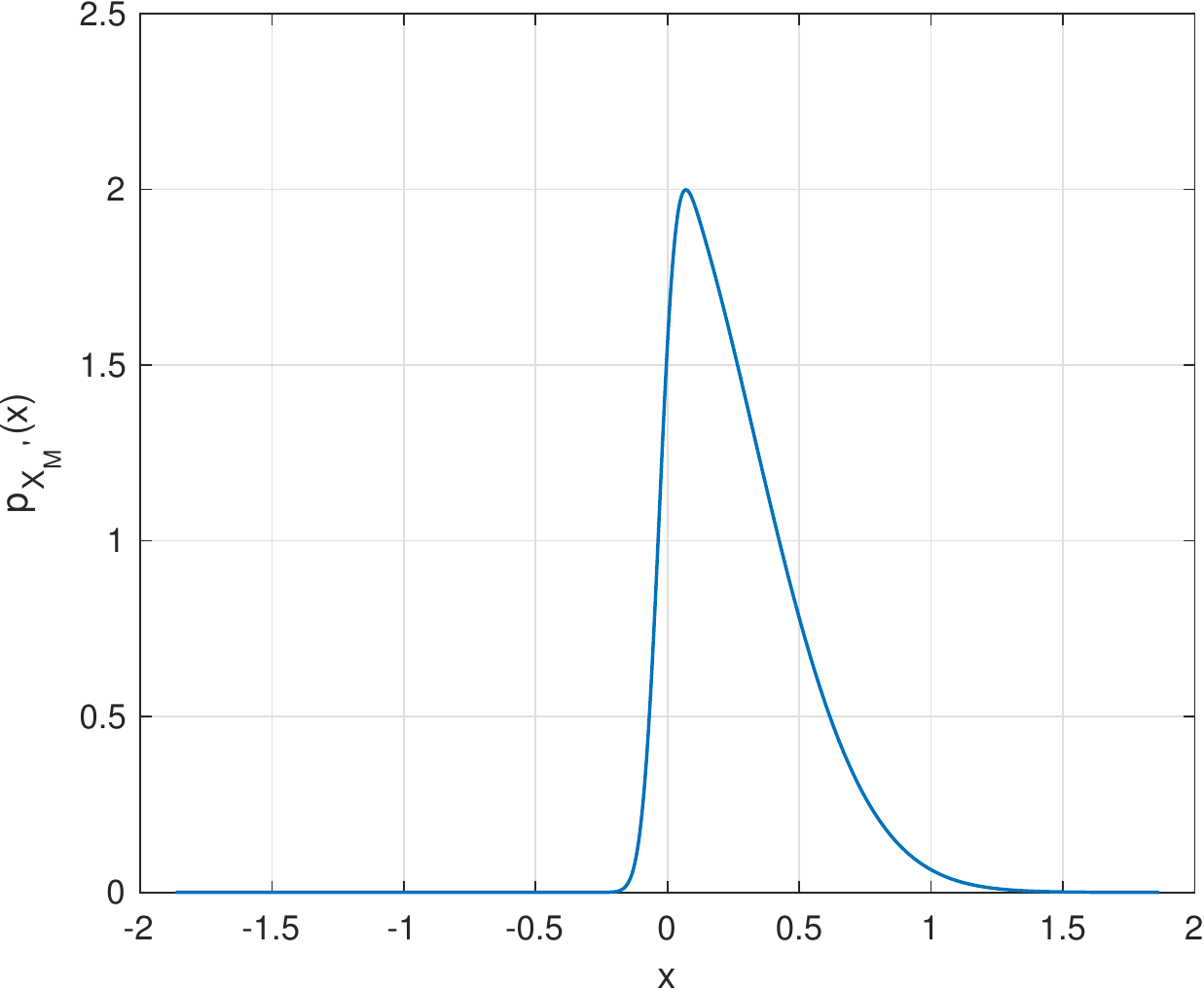}
\includegraphics[width=0.48\textwidth]{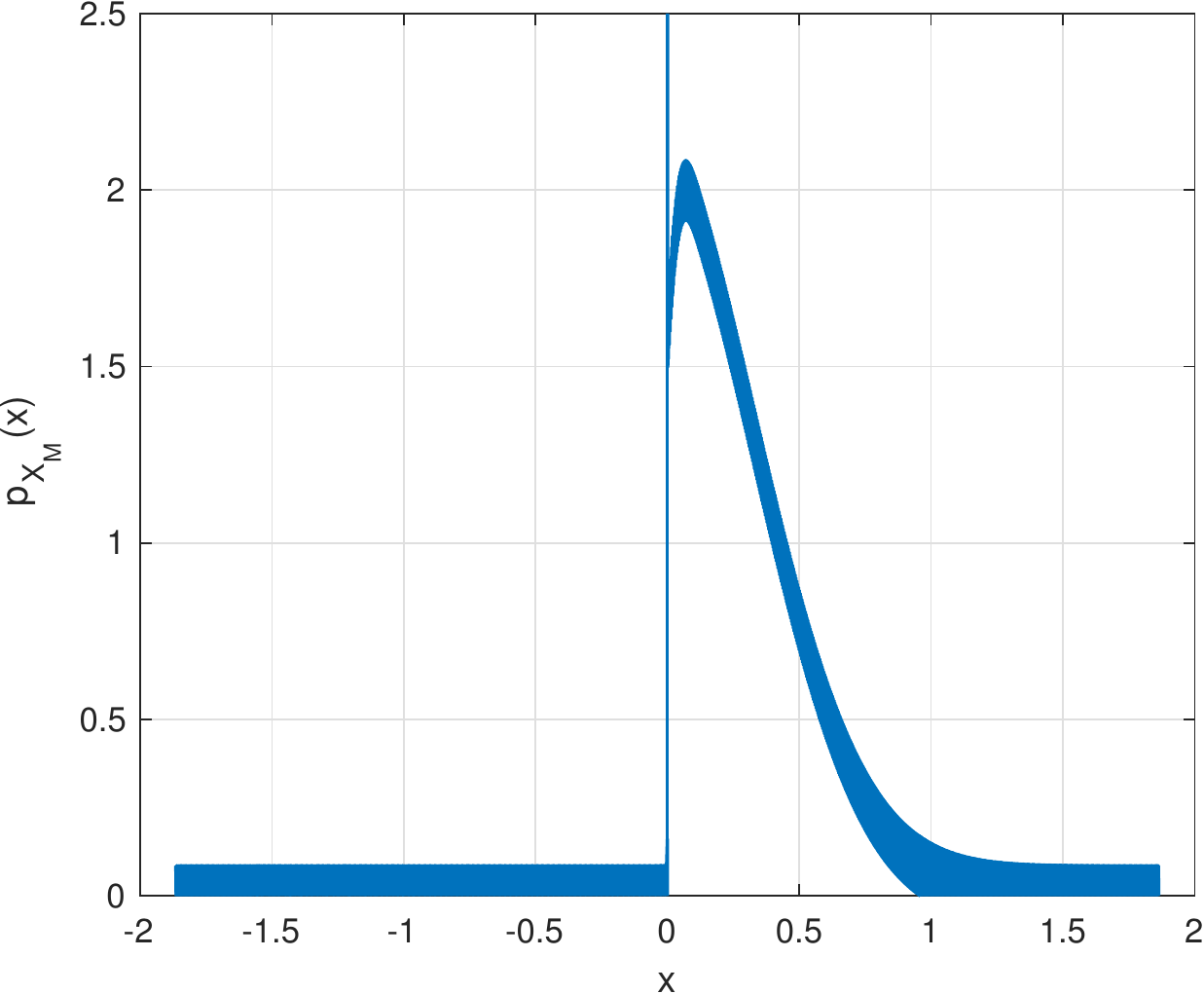}
\caption{PDF of the maximum of a discretely monitored Brownian motion for $t>0$ (left) and $t\geq0$ (right) for 50 monitoring dates with $\sigma$=0.4, risk-free interest rate $r=0.05$ and $2^{16}$ price grid points.}
\label{fig:7_qu_xmaxplot}
\end{center}
\vspace{-3mm}
\end{figure}

The calculation of the discretely monitored price for loopback options, as used by \cite{fusai2015}, are based on a distribution of the maximum (or minimum) of the process at $t>0$. Similarly to the schemes for barrier options also described in that paper, the first date is taken out of the Spitzer-based scheme and the result for $N-1$ dates is multiplied by the characteristic function. This gives a smooth probability distribution, i.e.\ $p_{X_{M'}}\in C^\infty$, as illustrated on the left hand plot of Figure \ref{fig:7_qu_xmaxplot}, where $X_{M'}$ is used to denote that we are using the maximum for $t>0$. As described by \cite{ruijter2015application}, Fourier based pricing methods using this PDF will therefore not be negatively affected by the Gibbs phenomenon and can achieve exponential error convergence with the number of log-price grid points. 

However, for the $\alpha$-quantile options we price in this paper, we require the distribution for the maximum (minimum) for $t\geq0$. As all L\'evy processes have the property that $X(0)=0$, the value of the maximum for $t\geq0$ cannot go below $0$. Therefore obtaining $p_{X_M}(x)$ using the Spitzer-based scheme with the full number of dates alters the PDF so that it now has an abrupt discontinuity and a large spike at $t=0$ as shown on the right hand panel of Figure \ref{fig:7_qu_xmaxplot}. The large spike corresponds to a single probability mass equal to $\int^0_{-\infty}p_{X_M'}(x)dx$. This can be understood by seeing that the cumulative distribution function (CDF) of $X_M$ is
\begin{equation}
P_{X_M}(x)=
\begin{cases}
0 & \text{for} \quad<0\\
P(X_M'\leq x) &\text{for} \quad x \geq 0.
\end{cases}
\end{equation}
If $P(X_M'\leq 0)\neq0$ as is generally the case for L\'evy processes, then $P_{X_M}(x)$ has a jump at $x=0$ of size $P(X_M'\leq 0)=\int^0_{-\infty}p_{X_M'}(x)dx$. Thus when the CDF is differentiated to give the PDF $p_{X_M}(x)$, this will contain a probability mass at $x=0$ of size $\int^0_{-\infty}p_{X_M'}(x)dx$.
We also note that the introduction of the discontinuity and spike has caused oscillations in the plot of the pdf via the Gibbs phenomenon. 

The existence of the discontinuity means that, as described by \cite{Boyd2001} and \cite{gottlieb1997gibbs} for example, we would no longer obtain exponential error convergence with grid size using these distributions to price options. We therefore use spectral filtering, as successfully implemented for Fourier based option-pricing methods by \cite{ruijter2015application}, \cite{Phelan2017} and \cite{cui2017equity}. Following it's use in previous literature we use an exponential filter of order 12 \citep{gottlieb1997gibbs}. 

\subsection{Results for discretely monitored $\alpha$-quantile options}
We implemented the pricing method described above using the numerical procedure described step-by-step in Appendix A.
Figure \ref{fig:quantileerrCPUonly} shows results with $\alpha=0.75$ for monitoring dates $N$ of 52, 252 and 1008 for the Gaussian, 
VG and Merton jump-diffusion processes. 

The error convergence for this method is extremely fast; we were able to achieve a CPU time of $10^{-2}$ seconds or less for an error of $10^{-8}$ using the Gaussian and Merton processes and $10^{-6}$ using the 
VG process.
There are too few points before reaching the error floor of $10^{-11}$, caused by the inverse $z$-transform, to assess whether the error convergence is exponential or high order polynomial. This is not a concern for using the method in practise but a more accurate $z$-transform would allow us to understand the convergence better.

In Table \ref{tab:7_quant_price} we show the variation of the price with $\alpha$ for 252 monitoring dates, as expected the option price increases with $\alpha$, and compare the results with those from Monte-Carlo pricing methods. We use two Monte-Carlo methods, the first simulates $N$ monitoring date paths and selects the $\alpha N^{\mathrm{th}}$ smallest value. The other method uses the same approach as \cite{Ballotta2001} by combining the Monte-Carlo simulation with the Dassios-Port-Wendel identity. Two independent paths of length $\alpha N$ and $(1-\alpha)N$ dates are simulated and the sum of their respective infimum and supremum are calculated to provide an estimate of the $\alpha$-quantile. Further details about this Monte-Carlo scheme are included in Appendix \ref{sec:App_MC}.
\begin{figure}
\begin{center}
\includegraphics[width=0.32\textwidth]{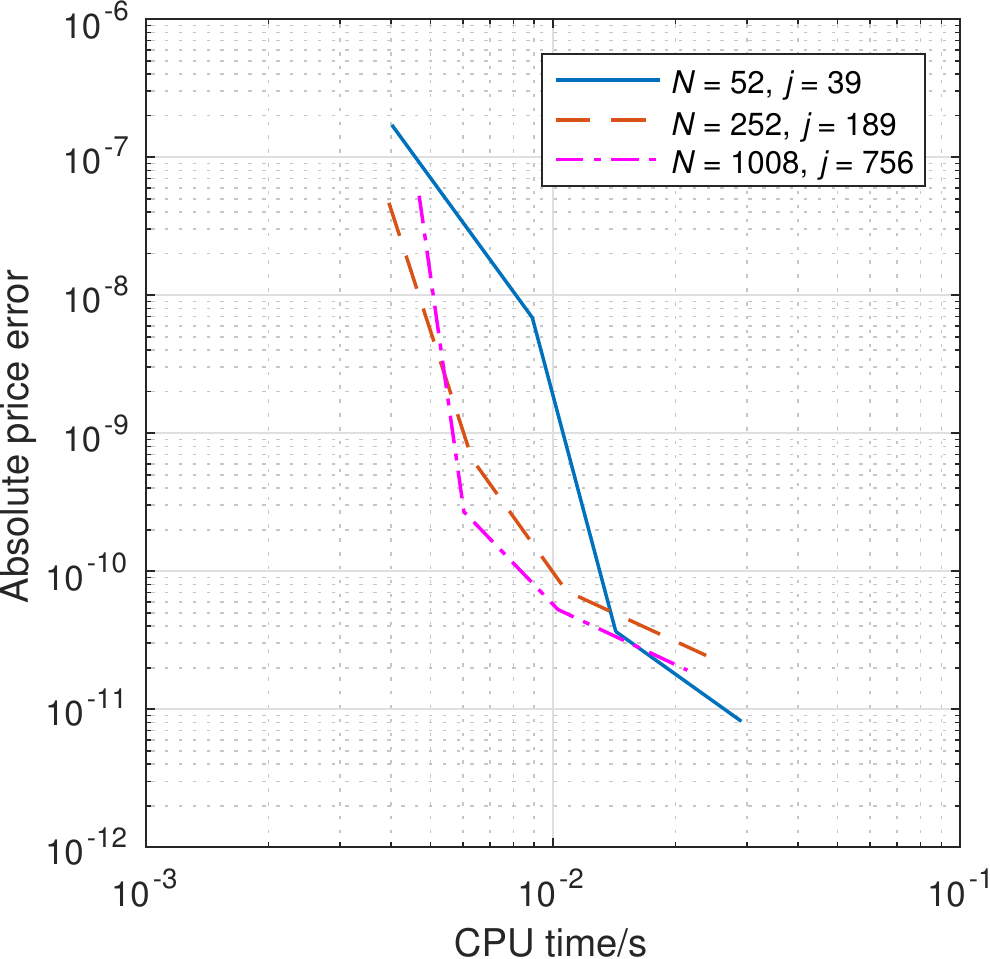}
\includegraphics[width=0.32\textwidth]{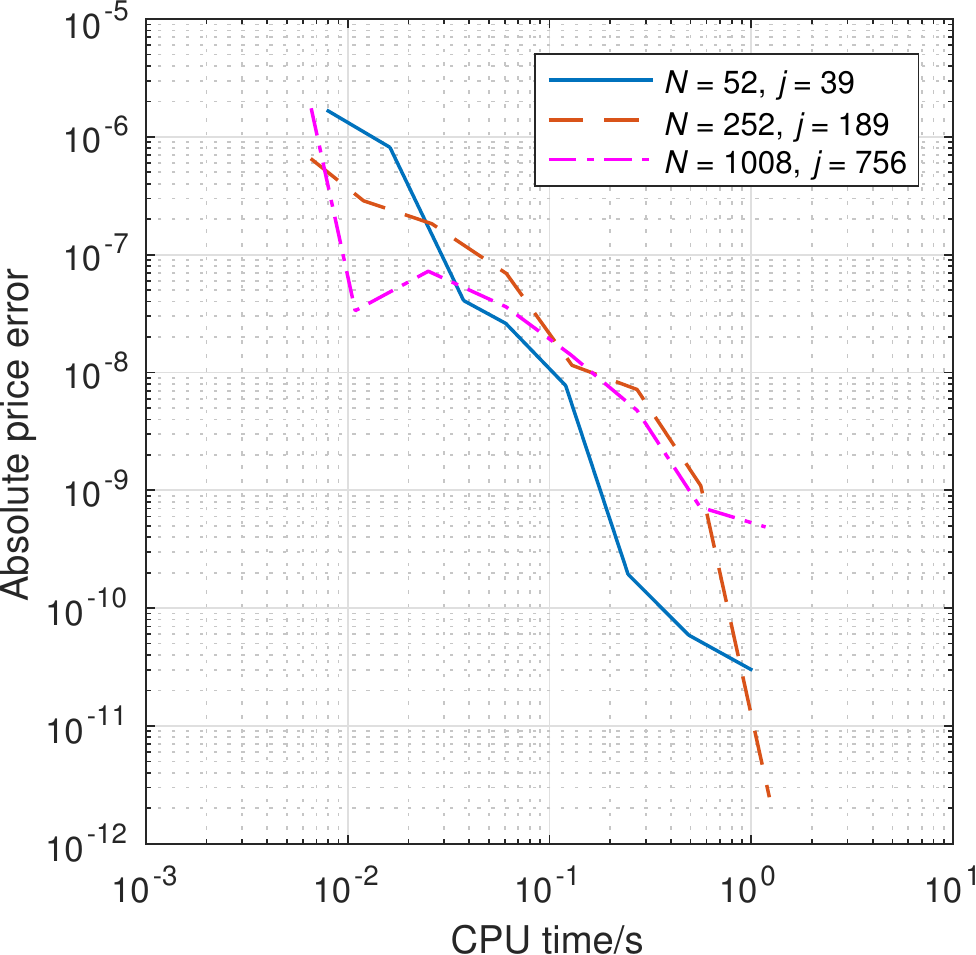}
\includegraphics[width=0.32\textwidth]{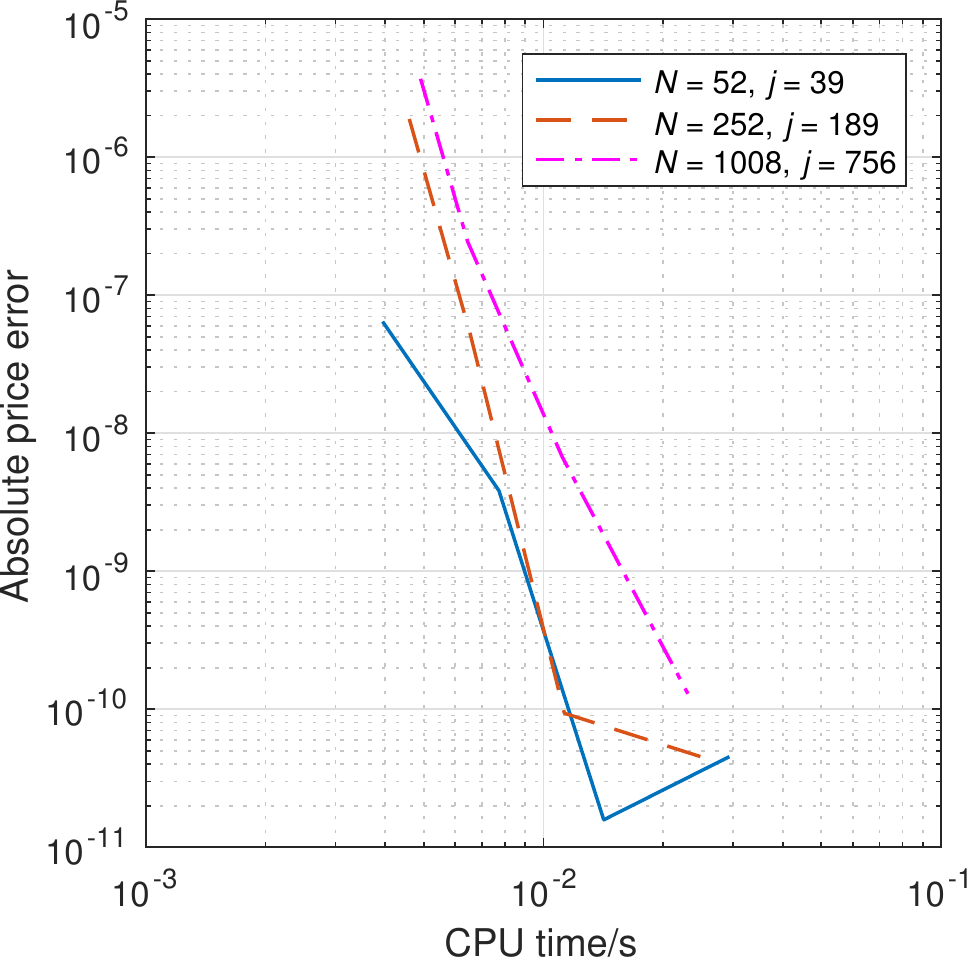}
\end{center}
\caption{Convergence of the pricing error with CPU time with an underlying asset driven by Gaussian (right), VG (centre) and Merton jump-diffusion processes. For the Gaussian and Merton jump-diffusion processes the log-price grid size range is $2^7-2^{12}$, and for the VG process the range is $2^8-2^{17}$.}
\label{fig:quantileerrCPUonly}
\end{figure}

\begin{table}[h] 
\centering
\begin{tabular}{r|r|rrr|rrr}
\hline
\hline
\multicolumn{8}{c}{Gaussian} \\
\hline
&Spitzer&\multicolumn{3}{c}{Monte Carlo}&\multicolumn{3}{|c}{Monte Carlo + DPW}\\
\hline
$\alpha$ & price & price & 2 std dev & difference& price &2 std dev & difference\\
\hline
$2/3$& 0.20847219 & 0.20862943 & 0.00044476 & -0.00015724 & 0.20874112 & 0.00046458 & 0.00026893\\
$3/4$& 0.24346198 & 0.24365156 & 0.00059366 & -0.00018958 & 0.24338425 & 0.00057616 & 7.773E-05\\
$5/6$& 0.28541002 & 0.2850389 & 0.00057114 & 0.00037112 & 0.28519695 & 0.00057528 & 0.00021307\\
\hline\hline
\multicolumn{8}{c}{VG} \\
\hline
&Spitzer&\multicolumn{3}{c}{Monte Carlo}&\multicolumn{3}{|c}{Monte Carlo + DPW}\\
\hline
$\alpha$ & price & price & 2 std dev & difference& price &2 std dev & difference\\
\hline
$2/3$& 0.15893822 & 0.15888004 & 0.000206 & 5.818E-05 & 0.15886432 & 0.00022294 & 7.39E-05\\
$3/4$& 0.17649028 & 0.17655861 & 0.00023774 & -6.833E-05 & 0.17658481 & 0.00022576 & -9.453E-05\\
$5/6$& 0.19604275 & 0.19588957 & 0.00024904 & 0.00015318 & 0.19595742 & 0.00024732 & 8.533E-05\\
\hline\hline
\multicolumn{8}{c}{Merton jump-diffusion} \\
\hline
&Spitzer &\multicolumn{3}{c}{Monte Carlo}&\multicolumn{3}{|c}{Monte Carlo + DPW}\\
\hline
$\alpha$ & price & price & 2 std dev & difference& price &2 std dev & difference\\
\hline
$2/3$& 0.15957161 & 0.15960486 & 0.0002189 & -3.325E-05 & 0.15970826 & 0.00023546 & -0.00013665\\
$3/4$&0.17757316 & 0.17761843 & 0.00026546 & -4.527E-05 & 0.17757363 & 0.00025204 & 4.7E-07\\
$5/6$ & 0.19776625 & 0.19778567 & 0.00026732 & -1.942E-05 & 0.19778468 & 0.00026808 & -1.843E-05\\
\hline
\hline
\end{tabular}
\caption{Comparison between the value of an $\alpha$-quantile option with 252 dates and $2^{16}$ price grid points compared with the value of the same contract using a Monte-Carlo approximation and a Monte-Carlo approximation combined with the Dassios-Port-Wendel identity. Notice that the prices calculated using the Spitzer-based method are within two standard deviations of the Monte-Carlo price.}
\label{tab:7_quant_price}
\end{table}

\FloatBarrier

\section{Perpetual Bermudan options} \label{sec:7_pbo}
Bermudan options have the same payoff as European options but they can be exercised at a discrete set of intermediate dates rather than simply at a final expiry date. They can also be thought of as a discrete version of American options and, indeed, the prices for Bermudan options are often used as a approximation for the value of American options, see e.g. \cite{Feng2013}. Perpetual Bermudan and American options have no expiry date and are therefore ``live" until they are exercised. Pricing perpetual options is an easier problem than the valuation of those with finite expiry as the infinite time horizon means the optimal exercise boundary is constant rather than a function of time. Indeed, closed-form formulas exist for perpetual American options with simple processes, whereas there are none for finite expiry options.

We look at two methods for pricing perpetual Bermudan options. Firstly we implement the method by \cite{GreenThesis2009} which uses residue calculus. We also implement a new method which uses Spitzer identities and show a novel way to calculate the optimal exercise barrier. New numerical truncation bounds for the log-price domain are specified; these are required due to the infinite time horizon.

\begin{figure}
\begin{center}
\includegraphics[width=\textwidth]{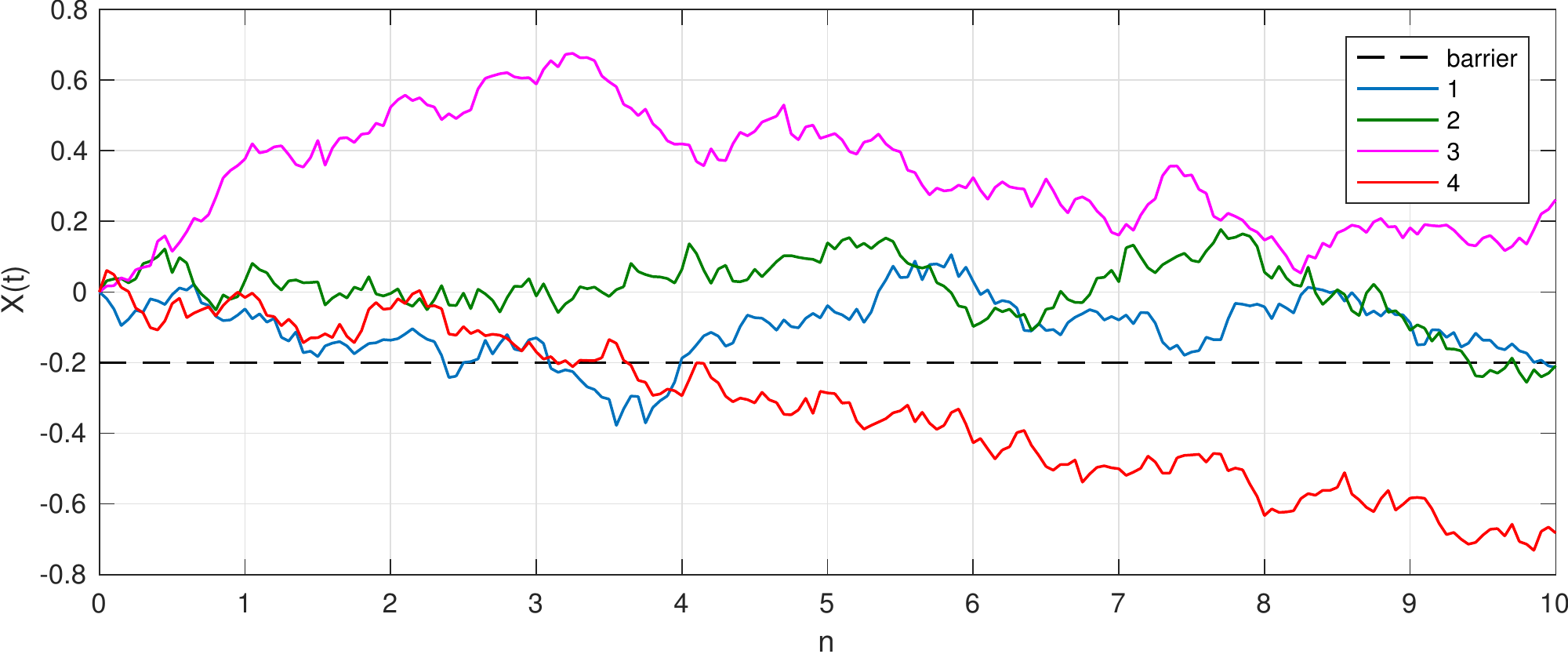}
\caption{Examples of discretely monitored continuous random paths with 10 monitoring dates. Notice that only paths 1 and 2 would contribute to the calculated distribution.}
\label{fig:7_qu_samplepaths}
\end{center}
\vspace{-3mm}
\end{figure}

Both methods require the probability distribution of the current value of a process, subject to the path not having crossed a lower barrier $l$, i.e.
\begin{equation}
p_l(x,n)dx=P[X_n\in(x,x+dx]\,\cap\,\max_{j<n}X_j>l].\label{eq:2_an_pldef}
\end{equation}
Using techniques from complex analysis, \cite{GreenThesis2009} and \cite{Green2010} showed that this can be expressed using the Spitzer identities as
\begin{align}
\widetilde{\widehat{p_l}}(x,n)&=
\begin{cases}
\mathcal{Z}^{-1}_{q\rightarrow n}\left[\mathcal{F}^{-1}_{\xi\rightarrow x}\left[\frac{P_{l+}(\xi,q)}{\Phi_\oplus(\xi,q)}\right]\right], & x\geq l,\\
\mathcal{Z}^{-1}_{q\rightarrow n}\left[\mathcal{F}^{-1}_{\xi\rightarrow x}\left[P_{l-}(\xi,q)\Phi_\ominus(\xi,q)\right]\right], & x< l.
\end{cases}\label{eq:lower_barrier}
\end{align}
For pricing single-barrier options, \cite{fusai2015} used the version of the Spitzer identity for $x\geq l$. In contrast, both the methods for pricing Bermudan options described here require the version of the identity for $x<l$. That is we require the distribution of $X(t_n)$ subject to $t_n$ being the first time it has crossed the discretely monitored barrier $l$. This is illustrated in Figure \ref{fig:7_qu_samplepaths} for a process with $10$ monitoring dates and a barrier of $-0.2$. The value of $X(t_{10})$ for paths 1 and 2 would contribute to the distribution as the paths stay above $l$ for monitoring dates $0$--$9$ but go below $l$ at the $10^{\mathrm{th}}$ monitoring date. Note that path 1 is acceptable as, even though its path does go below $l$ between dates $3$ and $4$, it is above the barrier at the actual monitoring dates. Path 3 does not count towards the distribution as it is above $l$ at the $10^{\mathrm{th}}$ monitoring date and path 4 does not count as it goes below $l$ at an earlier monitoring date. 

\subsection{Green's residue method} \label{sec:7_pbo_grm}

We here briefly recap the method described by \cite{GreenThesis2009} to provide a background to the results from our numerical implementation and also because we re-use some of the same techniques in deriving the new Spitzer based method described in Section \ref{sec:7_pbo_nfs}.
The scheme is based on a combination of a first-touch option paying $K-D$, where $K$ is the strike, 
and an overshoot option which pays the difference between the barrier $D$ and the underlying asset $S(t)=e^{X(t)}$ the first time the barrier is crossed. 
A first-touch option requires the probability that the first time the underlying asset crosses the barrier $l$ is time $n \Delta t$, i.e. 
\begin{align}
P[\tau_l=n\Delta t]=\int^l_{-\infty}p_l(x,n)dx.\label{eq:7_an_firsttouch}
\end{align}
Substituting $p_l(x,n) :=\mathcal{Z}^{-1}_{q\rightarrow n}\left[\mathcal{F}^{-1}_{\xi\rightarrow x}\left[P_{l-}(\xi,q)\Phi_\ominus(\xi,q)\right]\right]$ from Eq.~(\ref{eq:lower_barrier}) into this and taking the expectation over all dates, \cite{GreenThesis2009} obtains the price of the option as 
\begin{align}
v_{\mathrm{F}}(0,0)=(K-D)\sum^\infty_{n=1}e^{-r\Delta t n}\mathcal{Z}^{-1}_{q\rightarrow n}P_{l-}(0,q)\Phi_\ominus(0,q) \label{eq:7_an_ftprice}
\end{align}
where $\Delta t$ is the time step between monitoring dates. The next insight by \cite{GreenThesis2009} is that the summation and inverse $z$-transform cancel each other if we use $q=e^{-r\Delta t}$ 
because the summation equates to a forward $z$-transform. Then the price for a first-touch option is
\begin{align}
v_{\mathrm{F}}(0,0)=(K-D)P_{l-}\left(0,e^{-r\Delta t}\right)\Phi_\ominus\left(0,e^{-r\Delta t}\right).\label{eq:7_an_ftprice2}
\end{align}

The value of the payoff of an overshoot option at time $n$ is the expected overshoot $D-S(t_n)$ conditional on $n$ being the first time that the underlying asset process falls below $l$, $\tau_l$ is
\begin{align}
E[e^{-r\tau_l}(D-S(\tau_l))|\tau_l=n\Delta t]=e^{-rn\Delta t }\int^l_{-\infty}(D-S_0e^x)p_l(x,n)dx. 
\end{align}
We can calculate the option value by using tower property to take the expectation over all discrete monitoring and substitute $p_l(x,n) :=\mathcal{Z}^{-1}_{q\rightarrow n}\left[\mathcal{F}^{-1}_{\xi\rightarrow x}\left[P_{l-}(\xi,q)\Phi_\ominus(\xi,q)\right]\right]$ dates to obtain 
\begin{align}
v_{\mathrm{O}}(0,0)&=\sum^\infty_{n=1}e^{-rn\Delta t }\int^l_{-\infty}(D-S_0e^x)\mathcal{Z}^{-1}_{q\rightarrow n}\left[\mathcal{F}^{-1}_{\xi\rightarrow x}\left[P_{l-}(\xi,q)\Phi_\ominus(\xi,q)\right]\right]dx\nonumber\\
&=\frac{1}{2\pi}\int^{+\infty}_{-\infty}\widehat{\phi}(-\xi)P_{l-}(\xi,e^{-r\Delta t})\Phi_\ominus(\xi,e^{-r\Delta t})(\xi,e^{-r\Delta t})dx.
\end{align}
The final line uses the Plancherel relation to move into the Fourier domain and we have $q=e^{-r\Delta t}$ to cancel the inverse $z$-transform cancel as before. Using Eq.~(\ref{eq:Payoff}) for $\widehat{\phi}(\xi)$ for a put option with $a=-\infty$, $b=l$, $k=l$, $\alpha_{\mathrm{d}}>0$ and solving via residue method
\begin{align}
v_{\mathrm{O}}(0,0)&=-\frac{D}{2\pi}\int^{+\infty}_{-\infty}\frac{P_{l-}(\xi,e^{-r\Delta t})\Phi_\ominus(\xi,e^{-r\Delta t})}{\xi^2+i\xi}dx \nonumber\\
&=DP_{l-}(0,e^{-r\Delta t})\Phi_\ominus(0,e^{-r\Delta t})-DP_{l-}(-i,e^{-r\Delta t})\Phi_\ominus(-i,e^{-r\Delta t}). \label{eq:7_an_osprice}
\end{align}

As the price of a perpetual Bermudan option $v_\mathrm{B}(0,0)$ is the sum of a first-touch option with payoff $(K-D)$ from Eq.~(\ref{eq:7_an_ftprice}) and the price of an overshoot option from Eq.~(\ref{eq:7_an_osprice}) then
\begin{align}
v_{\mathrm{B}}(0,0)=v_{\mathrm{F}}(0,0)+v_{\mathrm{O}}(0,0)=KP_{l-}(0,e^{-r\Delta t})\Phi_\ominus(0,e^{-r\Delta t})-DP_{l-}(-i,e^{-r\Delta t})\Phi_\ominus(-i,e^{-r\Delta t}). \label{eq:7_an_pbprice}
\end{align}
\cite{GreenThesis2009} showed that, as the optimal exercise barrier gives a maximum price, solving $\partial v_{\mathrm{B}}(0,0)/\partial D=0$ for $D$ gives the optimal exercise barrier
\begin{align}
D_{\mathrm{opt}}=K\frac{\Phi_\ominus(0,e^{-r\Delta t})}{\Phi_\ominus(-i,e^{-r\Delta t})}.\label{eq:7_an_pbbarr}
\end{align}
Then, where $l_{\mathrm{opt}}=\log(D_{\mathrm{opt}}/S_0)$, we obtain the price of a perpetual Bermudan option
\begin{align}
v_{\mathrm{B}}(0,0)=K\Phi_\ominus(0,e^{-r\Delta t})[P_{l_{\mathrm{opt}}-}(0,e^{-r\Delta t})-P_{l_{\mathrm{opt}}-}(-i,e^{-r\Delta t})]. \label{eq:7_an_pbprice2}
\end{align}

\subsection{New formulation based on Spitzer identities} \label{sec:7_pbo_nfs}
The method by Green described in Section \ref{sec:7_pbo_grm} is very elegant mathematically but does not particularly lend itself to an intuitive understanding. Therefore we also devised an alternate method for pricing perpetual Bermudan options, including a new way of calculating the optimal exercise boundary.

We first recognise that the expected return from exercising a perpetual Bermudan put (subject to us being below the optimal exercise barrier) at monitoring date $n$ can be expressed as
\begin{align}
E[e^{-r n\Delta t }(K-S(t_n))^+]=e^{-rn\Delta t }\int^l_{-\infty}(K-S_0e^x)p_{l}(x,n)dx. \label{eq:7_an_pbsprice}
\end{align}
To obtain the value of the option we sum over all monitoring dates and we can also substitute $\widetilde{\widehat{p_{l}}}(\xi,q) :=P_{l-}(\xi,q)\Phi_\ominus(\xi,q)$ for the probability distribution $p_{l}(x,n)$
\begin{align}
v_{\mathrm{B}}(0,0)=\sum_{n=1}^{\infty}e^{-rn\Delta t }\int^l_{-\infty}(K-S_0e^x)\mathcal{Z}^{-1}_{q\rightarrow n}\mathcal{F}_{\xi\rightarrow x}^{-1}\widetilde{\widehat{p_{l}}}(\xi,q)dx. 
\end{align}
We can again use the trick by \cite{GreenThesis2009} of using $q=e^{-r\Delta t}$ so that the summation cancels with the inverse $z$-transform to give
\begin{align}
v_{\mathrm{B}}(0,0)=\int^l_{-\infty}(K-S_0e^x)\mathcal{F}_{\xi\rightarrow x}^{-1}\widetilde{\widehat{p_{l}}}(\xi,e^{-r\Delta t})dx. \label{eq:7_an_pbsprice2}
\end{align}
This integral can then be expressed in the Fourier domain using the Plancherel relation
\begin{align}
v_{\mathrm{B}}(0,0)&=\frac{1}{2\pi}\int^{+\infty}_{-\infty}\widehat{\phi}^*(\xi)\widetilde{\widehat{p_{l}}}(\xi,e^{-r\Delta t})dx
= \mathcal{F}^{-1}_{\xi\rightarrow x}\left[\widehat{\phi}^*(\xi)\widetilde{\widehat{p_{l}}}(\xi,e^{-r\Delta t})\right](0),\label{eq:7_an_pbsprice3}
\end{align}
where $\phi(\xi)$ is the damped payoff for a put option from Eq.~(\ref{eq:damped_payoff}) with $l$ being the optimal exercise boundary in the log-price domain.

\begin{figure}
\begin{center}
\includegraphics[width=\textwidth]{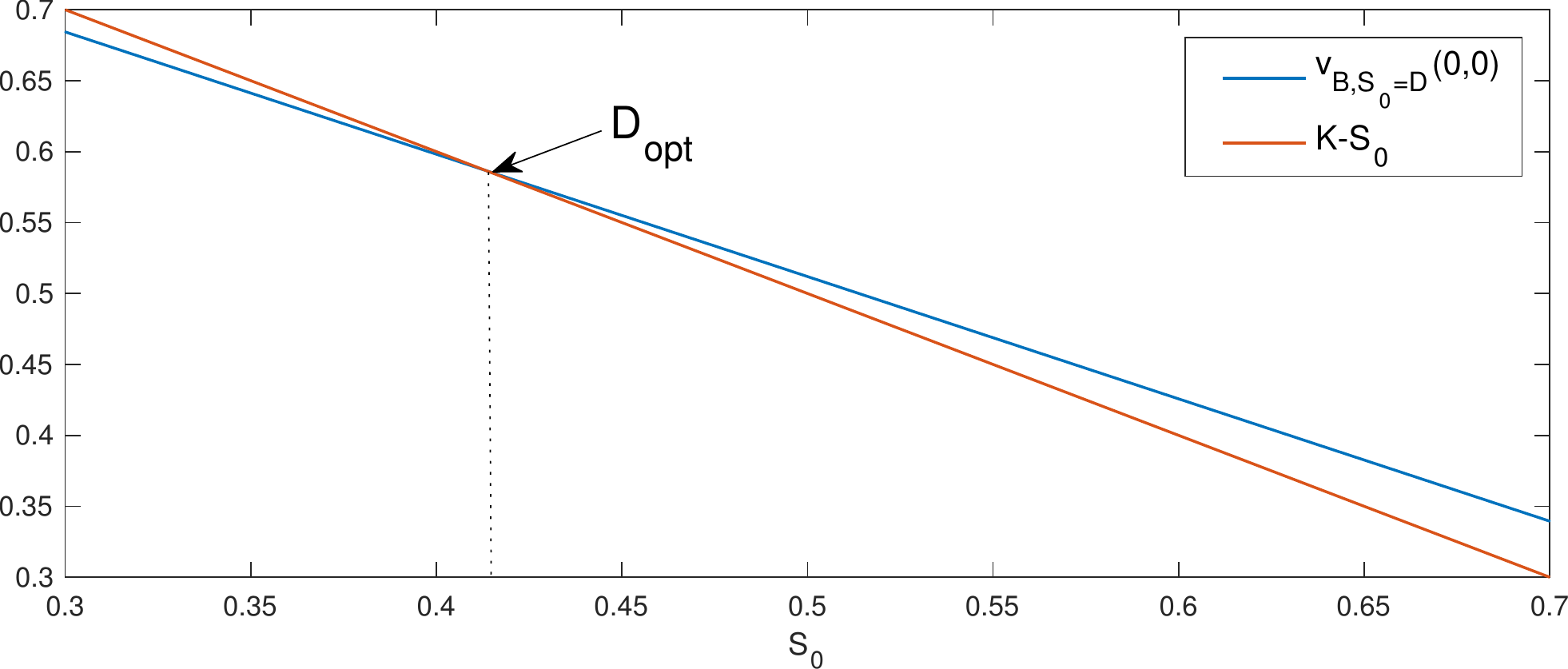}
\caption{Crossing point of the $K-S_0$ and $v_{\mathrm{B},S_0=D}(0,0)$ lines used to calculate the optimal exercise boundary $D_{\mathrm{opt}}$.}
\label{fig:7_pbs_Dcalgraph}
\end{center}
\vspace{-3mm}
\end{figure}

For this method, the calculation of the optimal exercise boundary is based on the idea that if the underlying asset is exactly at the optimal exercise boundary, i.e.\ $S_0=S(0)=D_{\mathrm{opt}}$, then the value of the payoff from exercising the option is equal to the expected value from continuing to hold the option. Furthermore, the boundary used to calculate $p_{l}(x,n)$ via $\widetilde{\widehat{p_{l}}}(\xi,q) :=P_{l-}(\xi,q)\Phi_\ominus(\xi,q)$ is $l=\log(D_{\mathrm{opt}}/S_0))$ and so $l=0$ and therefore we can rewrite Eq.~(\ref{eq:7_an_pbsprice}) as
\begin{align}
v_{\mathrm{B},S_0=D}(0,0)&=\int^0_{-\infty}(K-S_0e^{x})\mathcal{F}_{\xi\rightarrow x}^{-1}P_{0-}(\xi,e^{-r\Delta t})\Phi_\ominus(\xi,e^{-r\Delta t})dx\nonumber\\
&=\mathcal{F}_{\xi\rightarrow x}^{-1}\left[\widehat{\phi}^*(\xi)P_{0-}(\xi,e^{-r\Delta t})\Phi_\ominus(\xi,e^{-r\Delta t})\right](0).\label{eq:7_an_pbswrtd}
\end{align}
If we differentiate the expression on the first line of Eq.~(\ref{eq:7_an_pbswrtd}) with respect to $S_0$, we obtain
\begin{align}
\frac{\partial v_{\mathrm{B},S_0=D}(0,0)}{\partial S_0}=-\int^0_{-\infty}e^{x}\mathcal{F}_{\xi\rightarrow x}^{-1}P_{0-}(\xi,e^{-r\Delta t})\Phi_\ominus(\xi,e^{-r\Delta t})dx \label{eq:7_an_pbswrtdderiv},
\end{align}
which is constant. Therefore, if we were to plot Eq.~(\ref{eq:7_an_pbswrtd}) against $S_0$ we obtain a straight line and the point where that line crosses the payoff function $(K-S_0)$ represents the optimal exercise barrier. This is illustrated in Figure \ref{fig:7_pbs_Dcalgraph}.
Therefore, by calculating Eq.~(\ref{eq:7_an_pbswrtd}) for two values of $S_0$ ($D_1$ and $D_2$) we can obtain the corresponding straight line equation with gradient $m$ and $y$-axis intercept $c$. We can then calculate the optimal exercise boundary $D_{\mathrm{opt}}$, corresponding to the point where the lines cross, as 
\begin{align}
D_{\mathrm{opt}}=\frac{K-c}{m+1}. \label{eq:7_pbs_Dcalcalc}
\end{align}
We can also speed up the computational time by noting that if we set $D_1=0$ in Eq.~(\ref{eq:7_an_pbswrtd}) we obtain the price
\begin{align}
v_{\mathrm{B},S_0=D_1}(0,0)=\int^l_{-\infty}K\mathcal{F}_{\xi\rightarrow x}^{-1}P_{0-}(\xi,e^{-r\Delta t})\Phi_\ominus(\xi,e^{-r\Delta t})dx=KP_{0-}(0,e^{-r\Delta t})\Phi_\ominus(0,e^{-r\Delta t}),\label{eq:7_an_pbswrtd2}
\end{align}
which means that we only need to perform the inverse Fourier transform for the other calibration point($D_2$). To avoid computational errors, rather than calculating the Spitzer identity with $l=0$ we select $l=l_\epsilon$, where $l_\epsilon$ is the log-price domain grid step size spacing. This value does not depend on $S_0$ and so the calculation of the gradient in Eq.~(\ref{eq:7_an_pbswrtdderiv}) still returns a constant. The option price is then calculated using Eq.~(\ref{eq:7_an_pbsprice3}) with $l=D_{\mathrm{opt}}/S_0$.
\subsection{Calculation of truncation limits}
For the Fourier based methods used for finite expiry boundary options, the range of the log-price domain grid for the numerical calculations was based on the cumulants of the distribution over a single time step and, for the parameters used by \cite{fusai2015} and \cite{Phelan2017}, were calculated as approximately $\pm2$. However, for perpetual options we must consider the shape of the probability distribution far in the future. This is especially true when the risk-free rate $r$ is low as the contribution from future dates is discounted away extremely slowly. Therefore, the truncation limits used for finite expiry options are far too narrow for this application. We base our calculation of the new limits on the idea that the we should truncate the log-price domain at the value where the discount factor means any distortion of the distribution of the underlying asset process will have negligible effect on the final price calculation. We select a value that we wish the error to be below, i.e.\ $10^{-\lambda}$, and calculate the time it will take for the discount factor to be below this value 
\begin{align}
T_{\mathrm{bound}}=\lambda\log(10)/r.
\end{align}
We then approximate the standard deviation of the underlying process at this time with 
\begin{align}
\sigma_{\mathrm{bound}}=\sigma\sqrt{T_{\mathrm{bound}}},
\end{align}
where $\sigma$ is the underlying process volatility, normalised to 1 year. The bounds of the log-price domain $[-b,b]$ are now given by
\begin{align}
b=6\sigma_{\mathrm{bound}}.
\end{align}
\subsection{Results for perpetual Bermudan options with the Gaussian process}\label{sec:7_perpBermRes}
\begin{figure}[h]
\begin{center}
\includegraphics[width=0.48\textwidth]{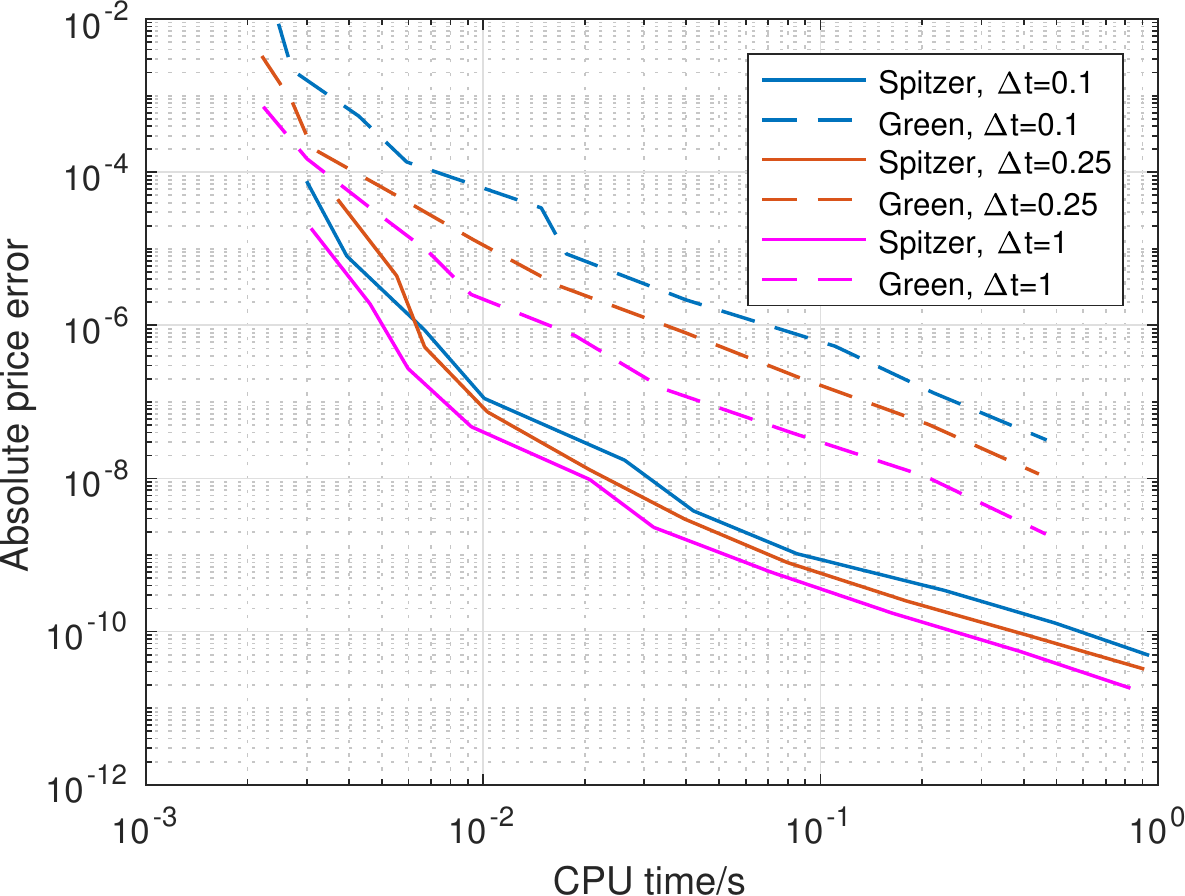}
\includegraphics[width=0.48\textwidth]{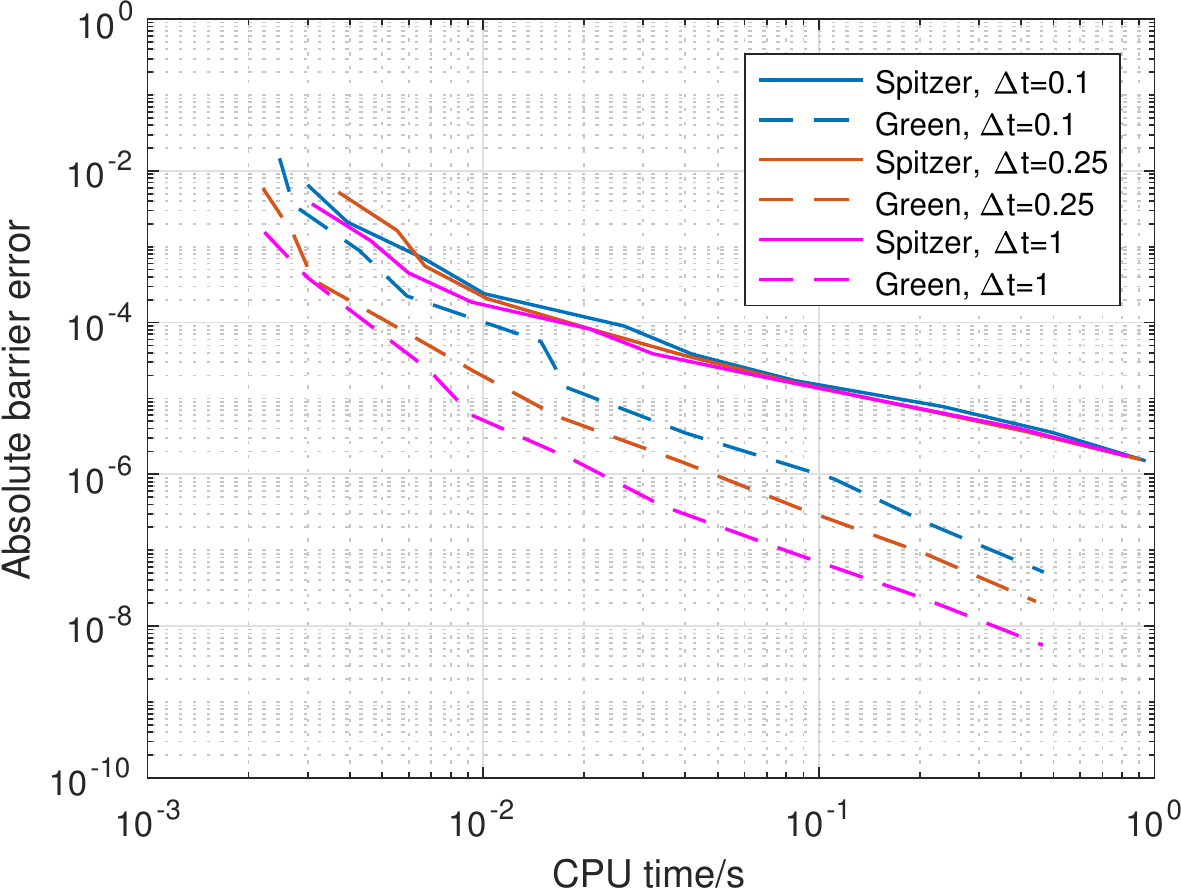}
\caption{Results for the price and barrier error convergence with CPU time with an underlying asset driven by a Gaussian process with a risk-free rate $r=0.02$. Notice that the pricing error convergence for the new method described in Section \ref{sec:7_pbo_nfs} labelled ``Spitzer" is faster than for the residue method described in Section \ref{sec:7_pbo_grm} labelled ``Green", whereas the barrier error convergence is worse.}
\label{fig:7_pb_perpberm_r_0_02_dt_all}
\end{center}
\end{figure}

\begin{figure}
\begin{center}
\includegraphics[width=0.48\textwidth]{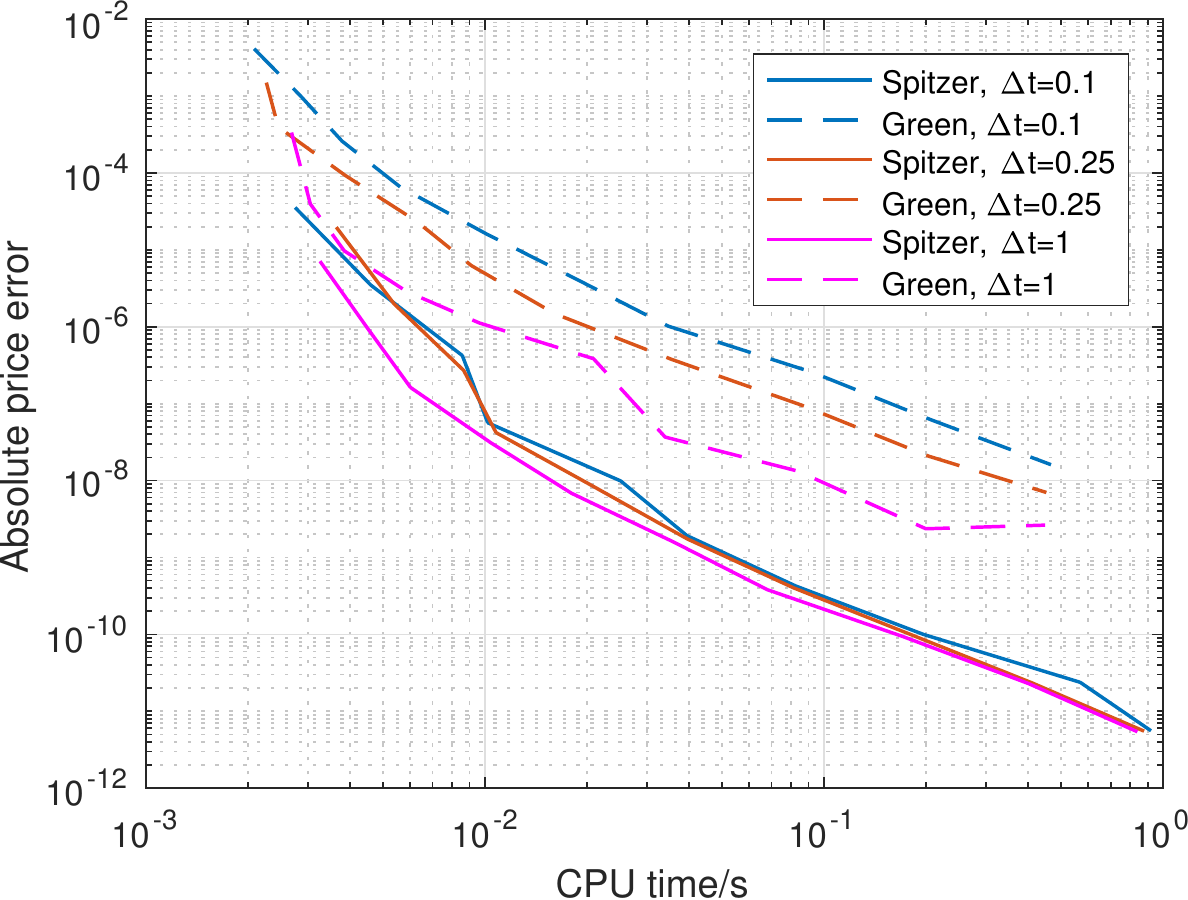}
\includegraphics[width=0.48\textwidth]{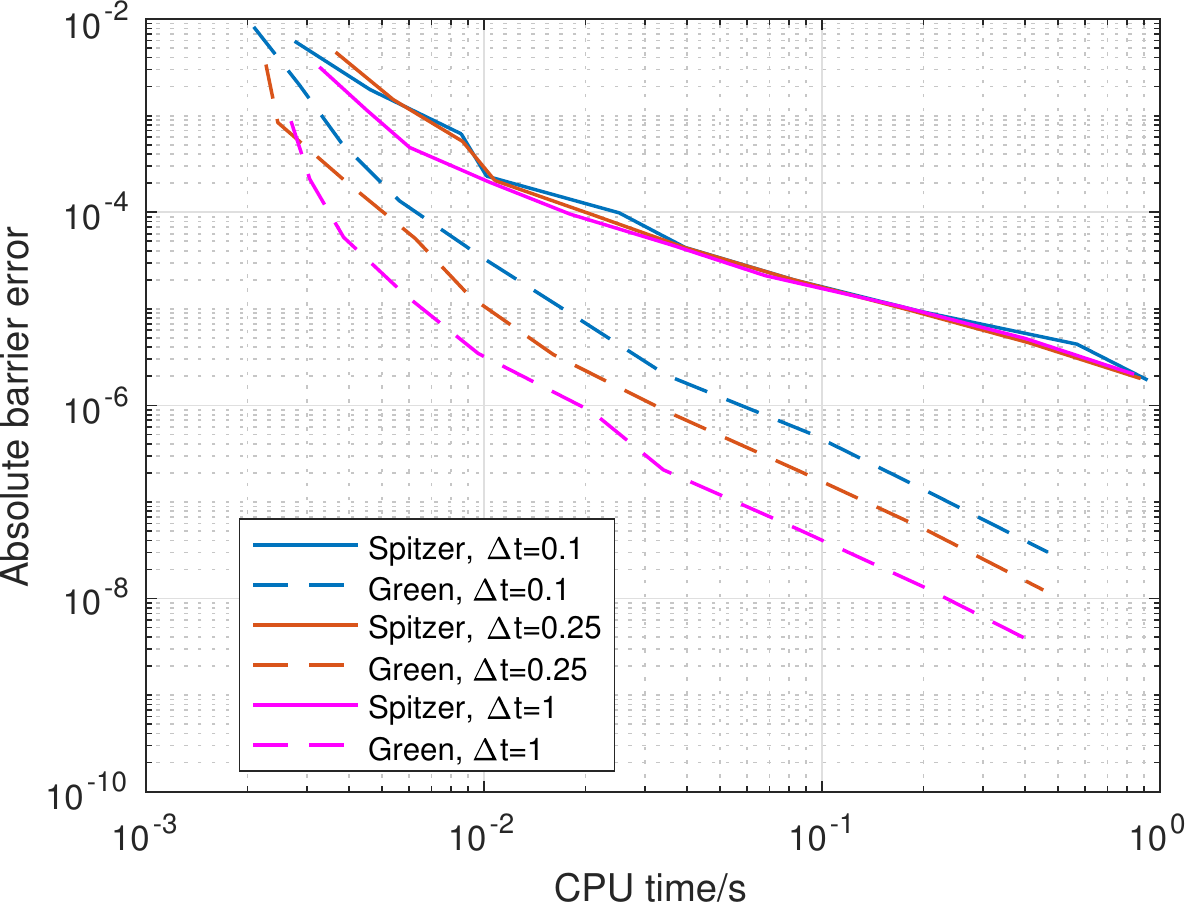}
\caption{Results for the price and barrier error convergence with CPU time with an underlying asset driven by a Gaussian process with a risk-free rate $r=0.05$. Notice that the pricing error convergence for the new method described in Section \ref{sec:7_pbo_nfs} labelled ``Spitzer" is faster than for the residue method described in Section \ref{sec:7_pbo_grm} labelled ``Green", whereas the barrier error convergence is worse.}
\label{fig:7_pb_perpberm_r_0_05_dt_all}
\end{center}
\end{figure}

Figures \ref{fig:7_pb_perpberm_r_0_02_dt_all} and \ref{fig:7_pb_perpberm_r_0_05_dt_all} show results for the two methods for pricing perpetual Bermudan options. The results labelled ``Green" are from the residue method described in Section \ref{sec:7_pbo_grm} and those labelled ``Spitzer" are from the new method described in Section \ref{sec:7_pbo_nfs}. Results are presented for error vs.\ CPU time for risk-free rates of 0.02 and 0.05. Using a Gaussian process for the underlying asset means the results as $\Delta t\rightarrow 0$ can be compared with closed-form calculations for perpetual American options. The convergence of the price for Bermudan options to the continuous case is shown in Tables \ref{tab:7_pbboth_price}--\ref{tab:7_pbboth_barrier} and 
we discuss the further verification of these results in Section \ref{sec:Res_other_Levy}.

Both methods perform well, and as $\Delta t\rightarrow 0$, the results approach those for the corresponding perpetual American option. The new Spitzer based method outperformed Green's residue method, achieving an error of $10^{-7}$ in approximately one tenth of the CPU time. In contrast, it is interesting to note that the convergence of the barrier is slower for the Spitzer based method than Green's residue method. Moreover the barriers of both methods converge at the same rate or slower than the price so we can see that the barrier error has a limited effect on the price error. This can be understood by considering Figure \ref{fig:7_pbs_Dcalgraph}. 
In this plot $(K-S_0)$ represents the value of exercising the option, whereas $v(0,0)$ represents the value keeping the option. Close to $D_{\mathrm{opt}}$ this difference is extremely small and so a minor error in $D_{\mathrm{opt}}$ has minimal effect on the price.

\begin{table}[h] 
\centering
\begin{tabular}{r|rrr|rrr}
\hline
\hline
& \multicolumn{3}{|c}{Spitzer}& \multicolumn{3}{|c}{Green}\\
\hline
$\Delta t$&$r=0.1$ & $r=0.05$ & $r=0.02$ &$r=0.1$ & $r=0.05$ & $r=0.02$ \\
\hline
1 & 0.20169919 & 0.33181098 & 0.53155442 & 0.20169919 & 0.33181098 & 0.53155362\\
0.5 & 0.20737414 & 0.33522271 & 0.53328533 & 0.20737414 & 0.33522271 & 0.53328453\\
0.25& 0.21021533 & 0.33695115 & 0.53414348& 0.21021533 & 0.33695115 & 0.53414268\\
0.1& 0.21197983 & 0.33798846 & 0.53465453 & 0.21197984 & 0.33798846 & 0.53465373\\
0.01 & 0.21305064 & 0.33861012 & 0.53495870 & 0.21305065 & 0.33861013 & 0.53495791\\
American & 0.21317038 & 0.33867902 & 0.53499224 & 0.21317038 & 0.33867902 & 0.53499224\\
\hline
\hline
\end{tabular}
\caption{Results for both methods for perpetual Bermudan options with $2^{20}$ price grid points showing the convergence to the price for a perpetual American option. }
\label{tab:7_pbboth_price}
\end{table}

\begin{table}[h] 
\centering
\begin{tabular}{r|rrr|rrr}
\hline
\hline
& \multicolumn{3}{|c}{Spitzer}& \multicolumn{3}{|c}{Green}\\
\hline
$\Delta t$&$r=0.1$ & $r=0.05$ & $r=0.02$ &$r=0.1$ & $r=0.05$ & $r=0.02$ \\
\hline
1 & 0.68360194 & 0.47916587 & 0.25110359 & 0.68360124 & 0.47916517 & 0.25110066\\
0.5 & 0.64678075 & 0.45056945 & 0.23520061 & 0.64678008 & 0.45056880 & 0.23519670\\
0.25 & 0.62026498 & 0.43075700 & 0.22442626 & 0.62026434 & 0.43075637 & 0.22442094\\
0.1& 0.59653531 & 0.41350275 & 0.21519240 & 0.59653469 & 0.41350214 & 0.21518427\\
0.01& 0.56851217 & 0.39363474 & 0.20472988 & 0.56851159 & 0.39363417 & 0.20470522\\
American & 0.55555556 & 0.38461538 & 0.20000000& 0.55555556 & 0.38461538 & 0.20000000 \\
\hline
\hline
\end{tabular}
\caption{Barrier calculated using both methods for perpetual Bermudan options with $2^{20}$ price grid points showing the convergence to the barrier for a perpetual American option.}
\label{tab:7_pbboth_barrier}
\end{table}

\subsection{Results for perpetual Bermudan options with other L\'evy processes}\label{sec:Res_other_Levy}
\begin{figure}[h]
\begin{center}
\includegraphics[width=0.48\textwidth]{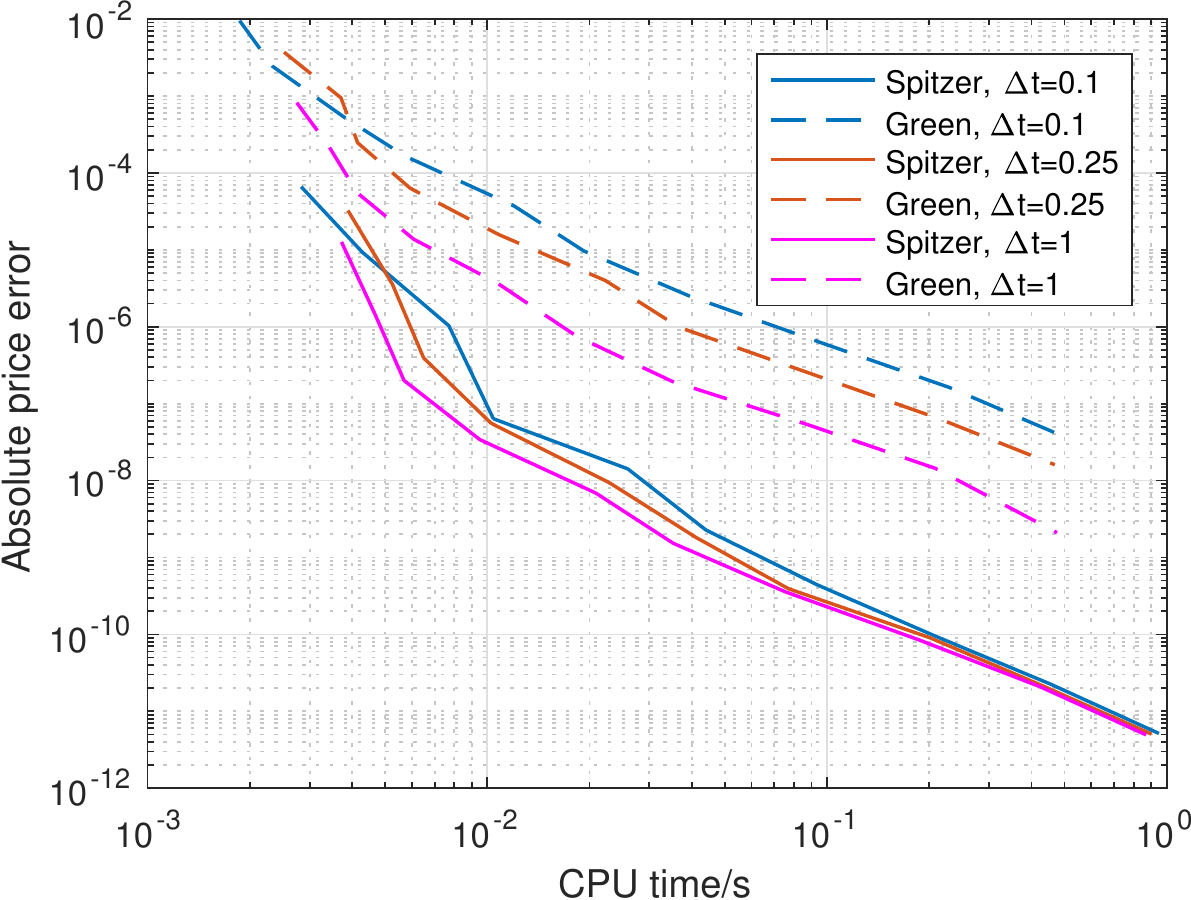}
\includegraphics[width=0.48\textwidth]{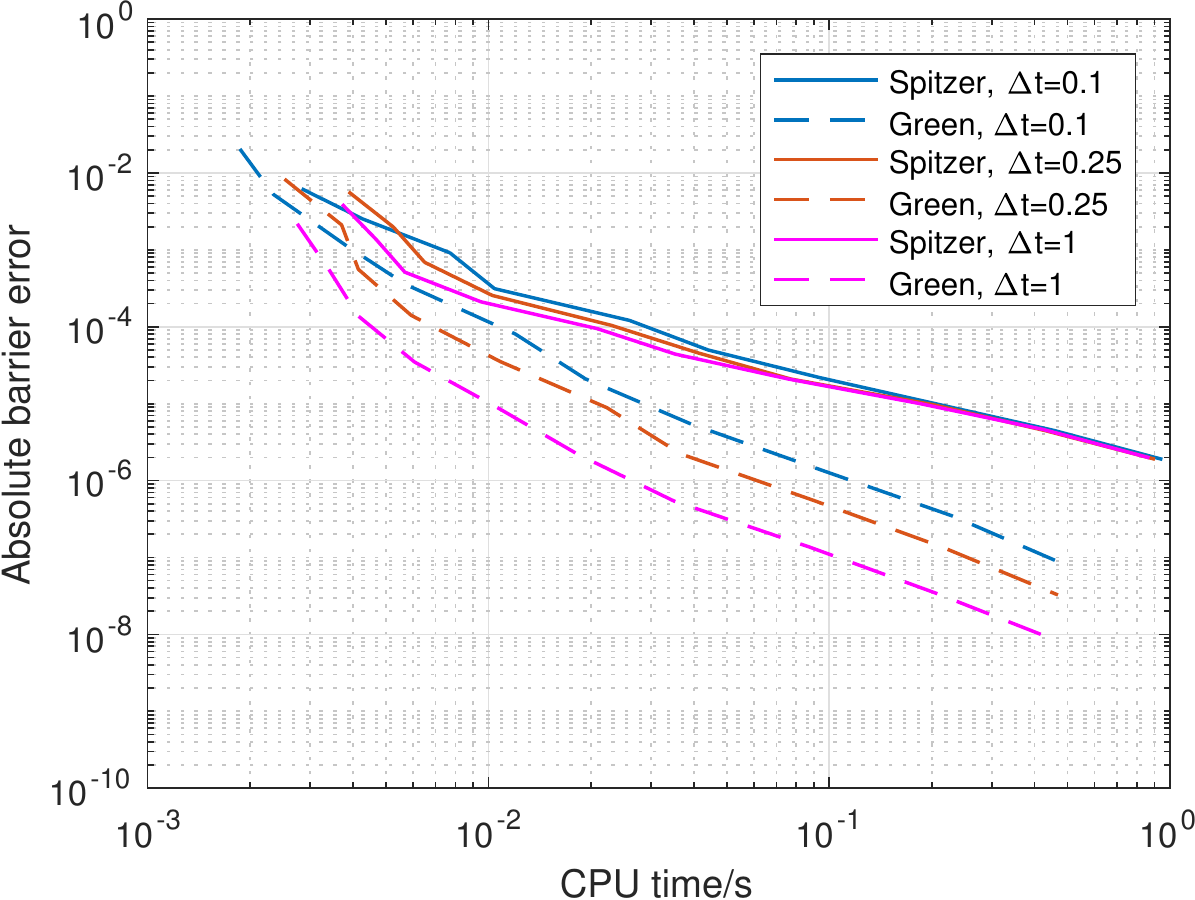}
\caption{Results for the price and barrier error convergence with CPU time with an underlying asset driven by a VG process with a risk-free rate $r=0.02$. Notice that the pricing error convergence for the new method described in Section \ref{sec:7_pbo_nfs} labelled ``Spitzer" is faster than for the residue method described in Section \ref{sec:7_pbo_grm} labelled ``Green", whereas the barrier error convergence is worse.}
\label{fig:7_pb_perpberm_r_0_02_dt_allVG}
\end{center}
\end{figure}

\begin{figure}[h]
\begin{center}
\includegraphics[width=0.48\textwidth]{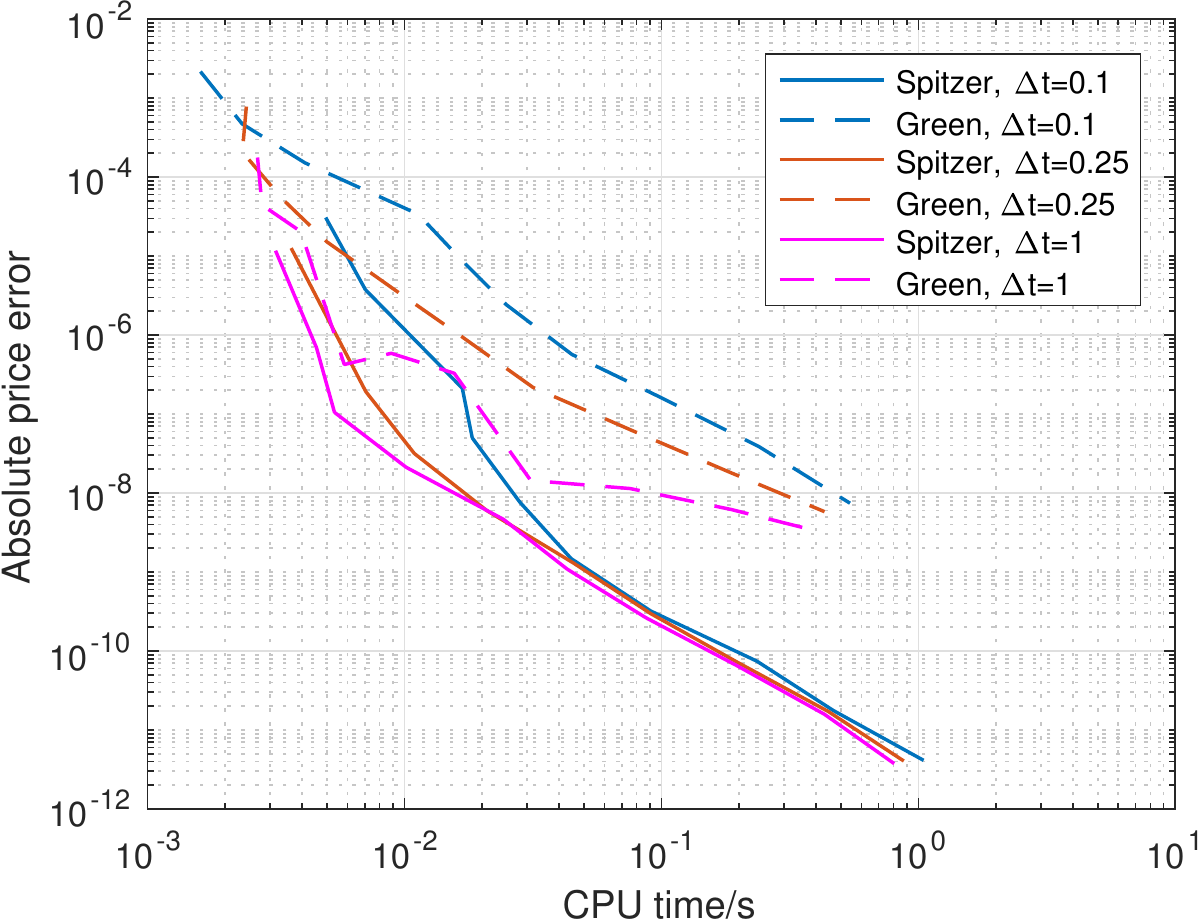}
\includegraphics[width=0.48\textwidth]{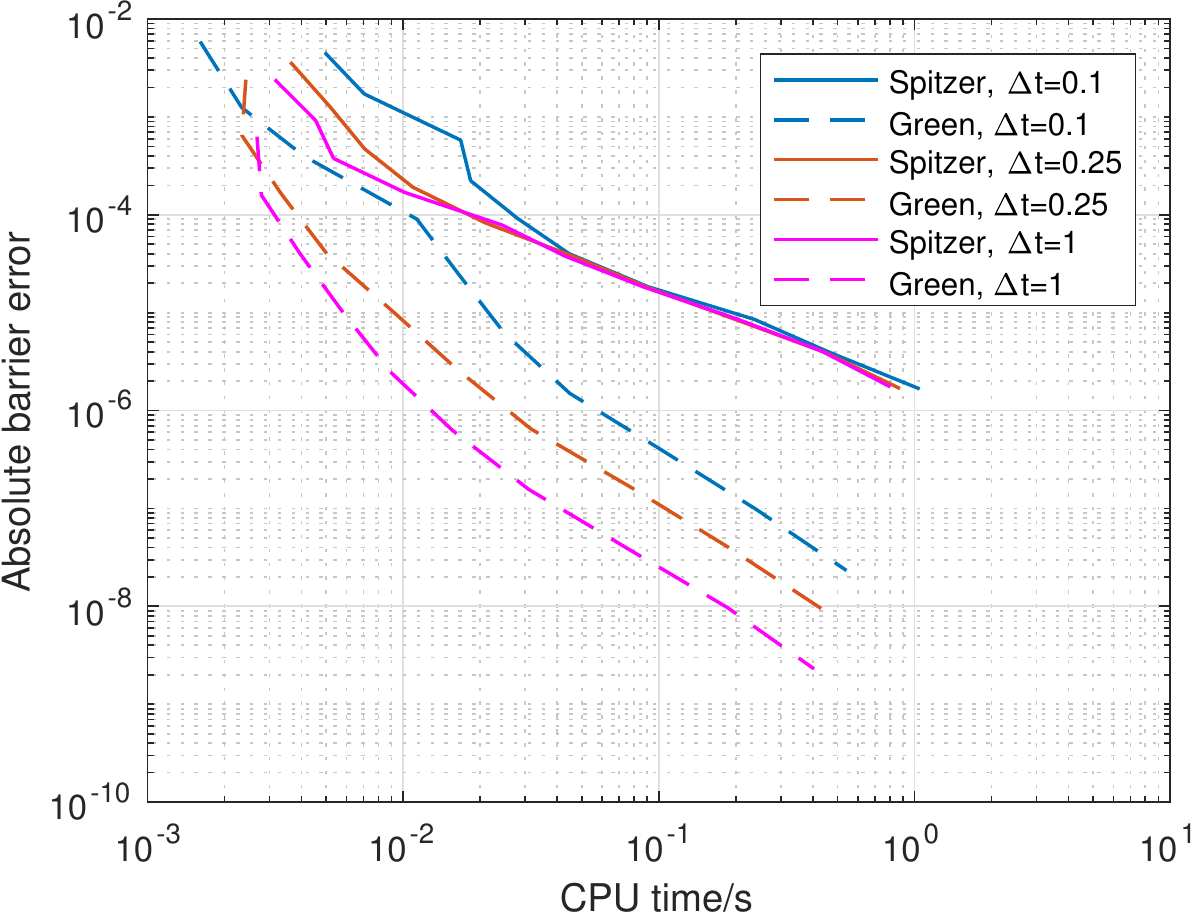}
\caption{Results for the price and barrier error convergence with CPU time with an underlying asset driven by a VG process with a risk-free rate $r=0.05$. Notice that the pricing error convergence for the new method described in Section \ref{sec:7_pbo_nfs} labelled ``Spitzer" is faster than for the residue method described in Section \ref{sec:7_pbo_grm} labelled ``Green", whereas the barrier error convergence is worse.}
\label{fig:7_pb_perpberm_r_0_05_dt_allVG}
\end{center}
\end{figure}

\begin{figure}[h]
\begin{center}
\includegraphics[width=0.48\textwidth]{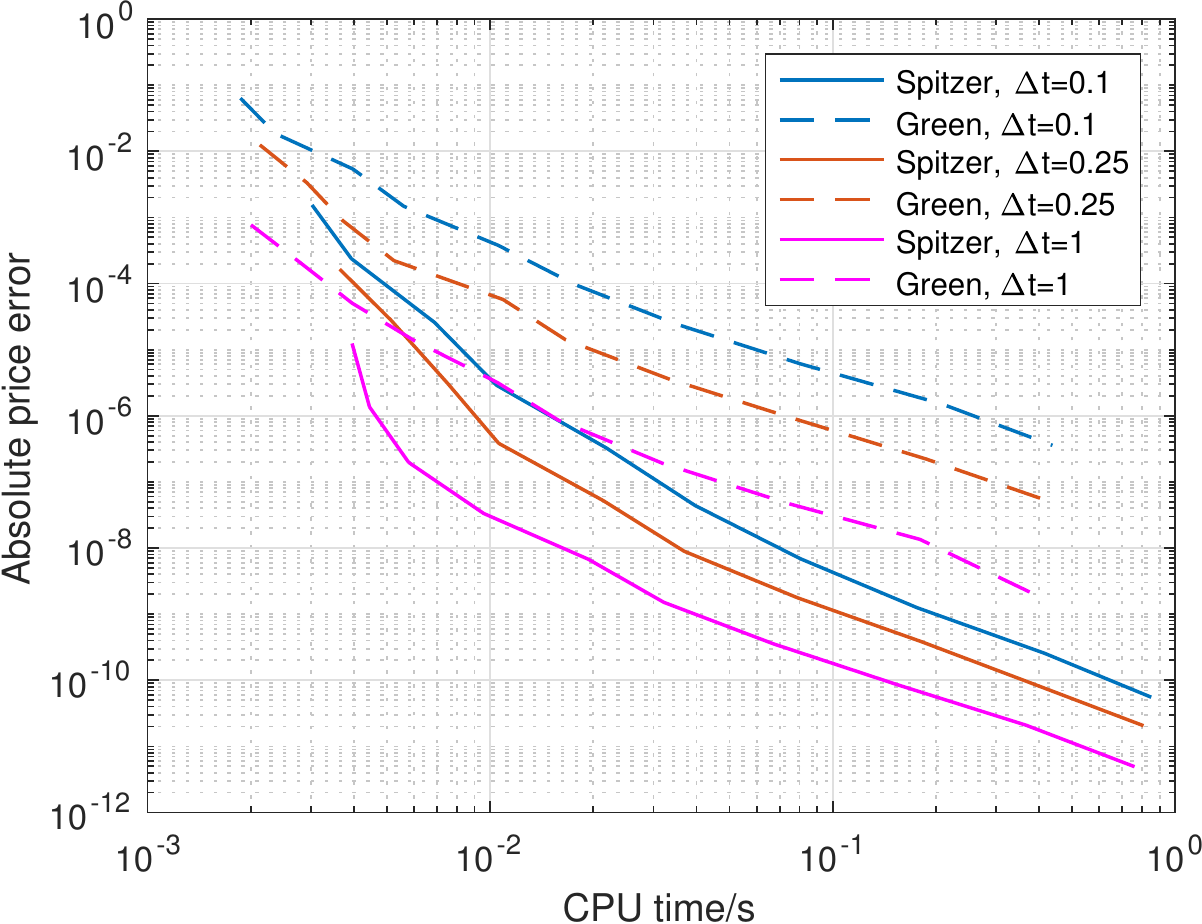}
\includegraphics[width=0.48\textwidth]{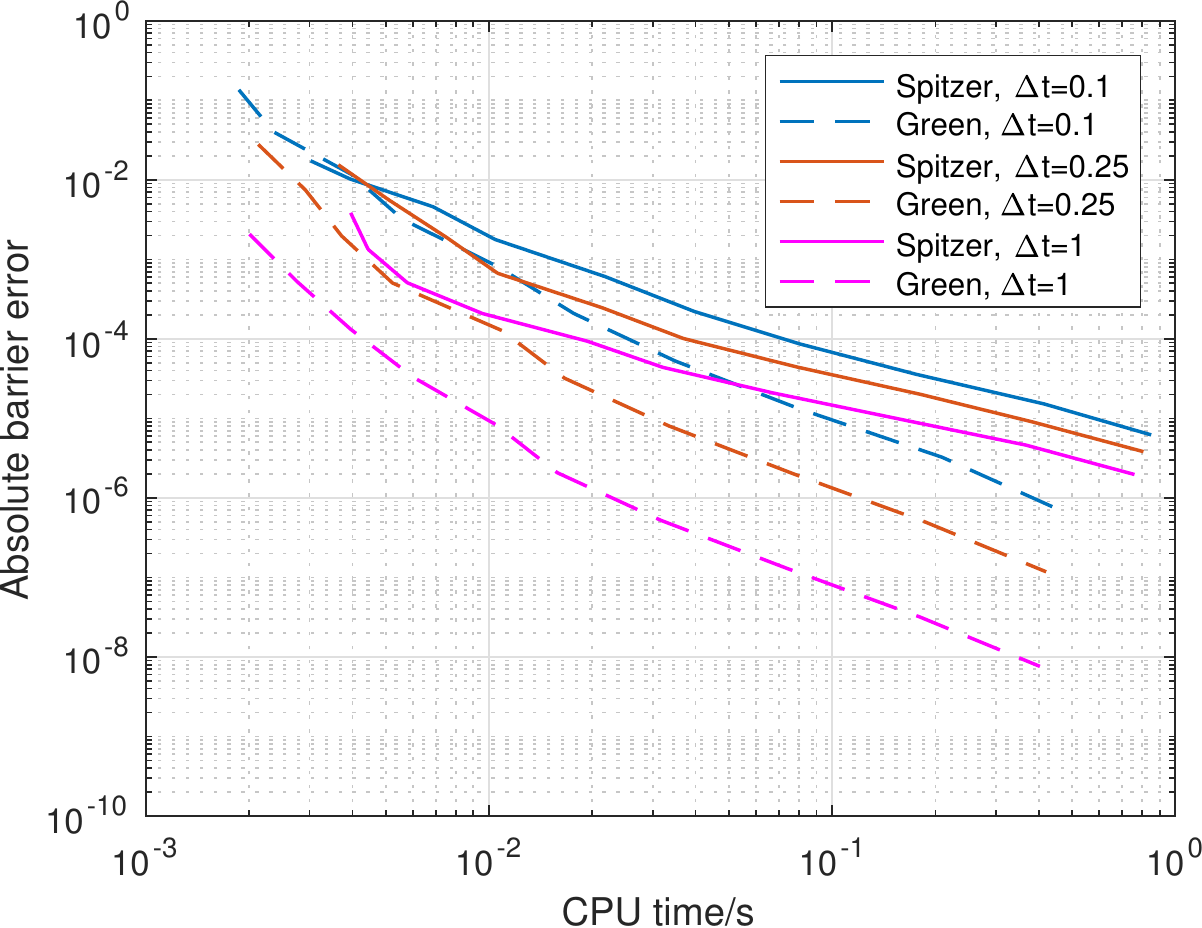}
\caption{Results for the price and barrier error convergence with CPU time with an underlying asset driven by a Merton process with a risk-free rate $r=0.02$. Notice that the pricing error convergence for the new method described in Section \ref{sec:7_pbo_nfs} labelled ``Spitzer" is faster than for the residue method described in Section \ref{sec:7_pbo_grm} labelled ``Green", whereas the barrier error convergence is worse.}
\label{fig:7_pb_perpberm_r_0_02_dt_allMerton}
\end{center}
\end{figure}

\begin{figure}[h]
\begin{center}
\includegraphics[width=0.48\textwidth]{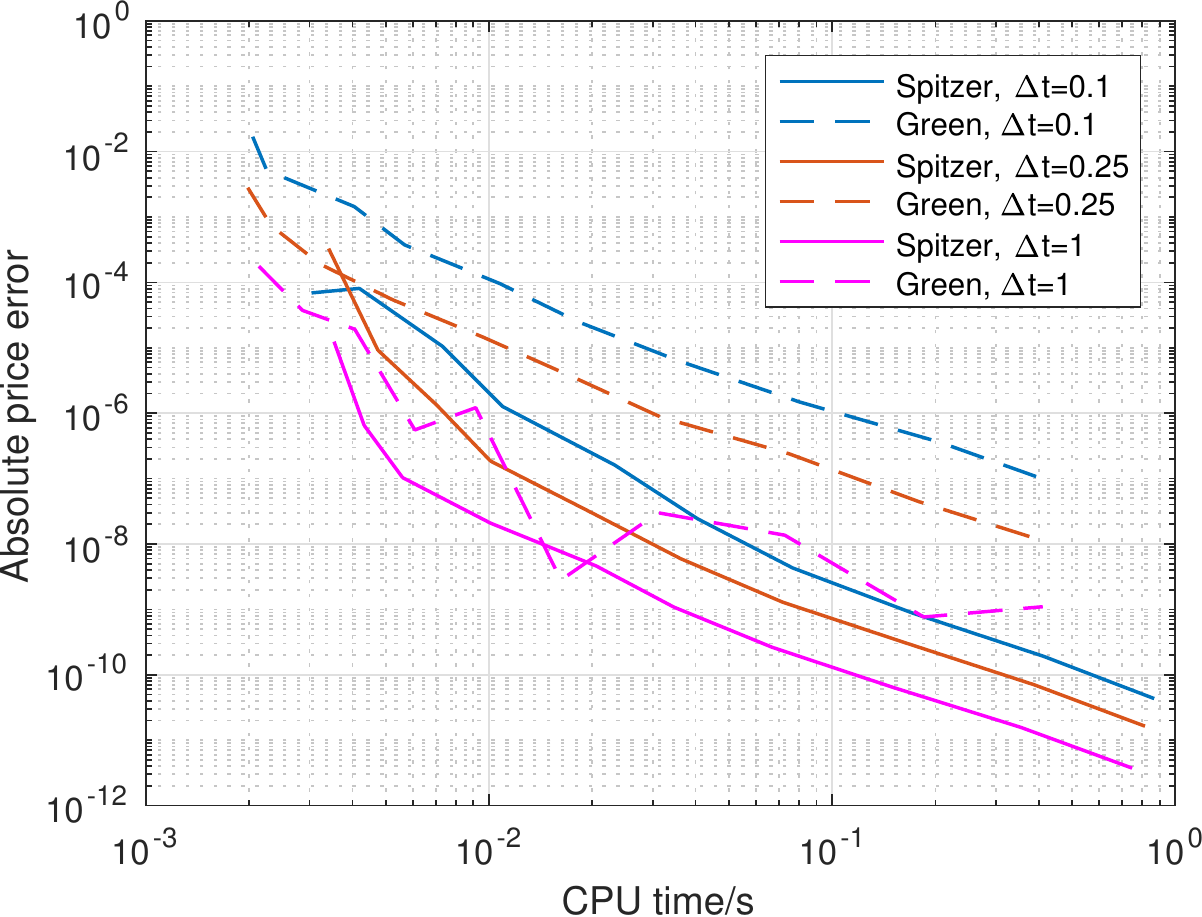}
\includegraphics[width=0.48\textwidth]{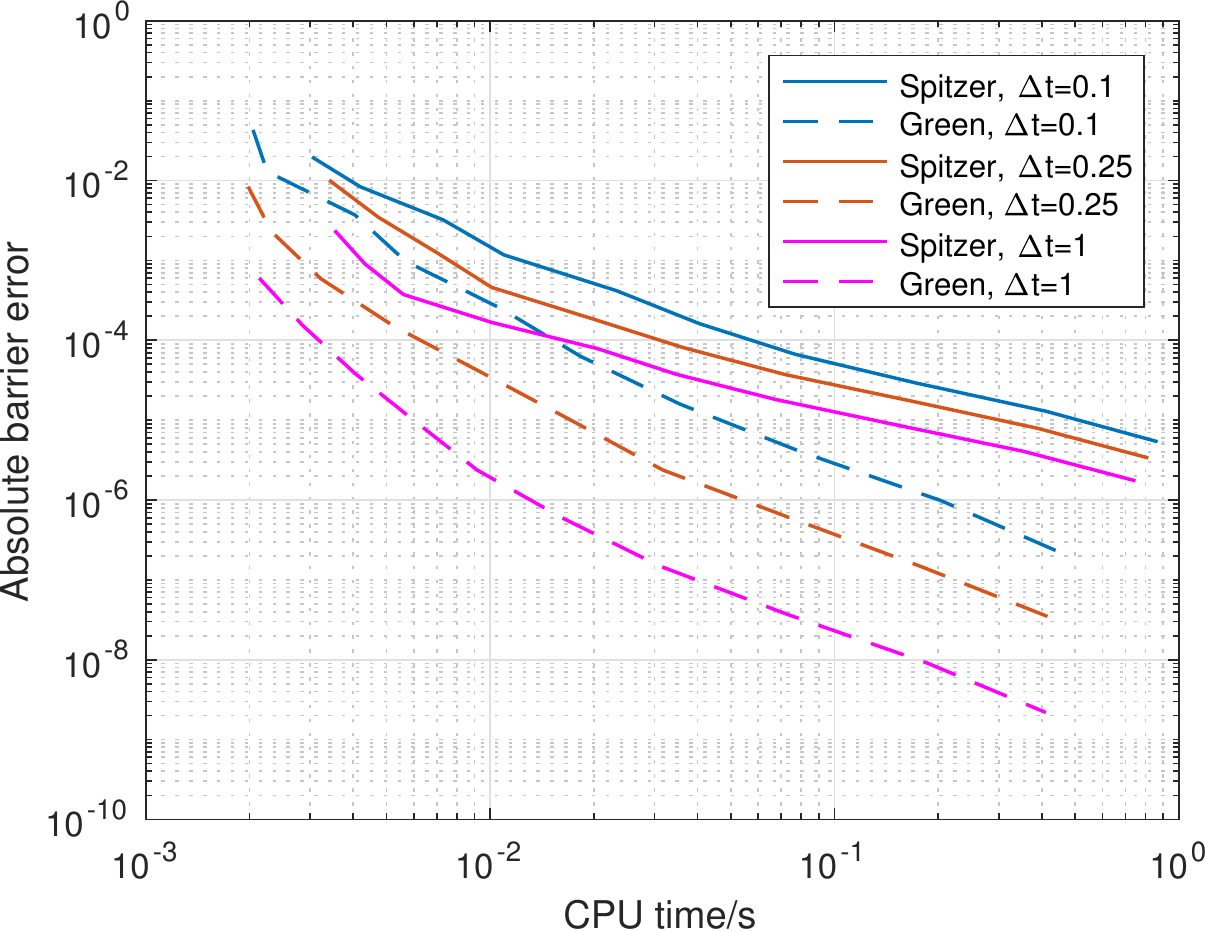}
\caption{Results for the price and barrier error convergence with CPU time with an underlying asset driven by a Merton process with a risk-free rate $r=0.05$. Notice that the pricing error convergence for the new method described in Section \ref{sec:7_pbo_nfs} labelled ``Spitzer" is faster than for the residue method described in Section \ref{sec:7_pbo_grm} labelled ``Green", whereas the barrier error convergence is worse.}
\label{fig:7_pb_perpberm_r_0_05_dt_allMerton}
\end{center}
\end{figure}

Figures \ref{fig:7_pb_perpberm_r_0_02_dt_allVG}--\ref{fig:7_pb_perpberm_r_0_05_dt_allMerton} show results for the price and optimal exercise barrier error vs.\ the number of grid points and CPU time, with the underlying asset driven by the VG and Merton processes. Once again, the Spitzer-based method performs better, achieving an error of $10^{-7}$ about 10 times quicker than Green's method. The error is calculated as the precision compared to the result with the maximum number of grid points and we discuss the verification of these results below.

In Tables \ref{tab:7_pbboth_price} and \ref{tab:7_pbboth_barrier} we showed that that the results with the Gaussian process converge to the closed-form solution by \cite{Merton1973} as $\Delta t\rightarrow 0$. However, we do not have a general closed-form solution for other L\'evy processes, so a Monte-Carlo method was used as an approximation.

We wrote Monte-Carlo pricing procedures with underlying assets driven by Gaussian, VG and Merton jump-diffusion processes. 
Although the discrete nature of perpetual Bermudan options is appropriate for a Monte-Carlo simulation, the absence of an expiry date means that a Monte-Carlo scheme with a finite number of dates will not be a true representation of the contract. However, in our simulation we truncate the Monte-Carlo simulation far enough in the future that the effect of disregarding these future dates is less than the standard deviation of the Monte-Carlo method itself (clearly this method is more feasible for high discount factors and large time steps, as the effect of future dates is discounted away more rapidly). The calculation of the optimal exercise barrier uses the same philosophy as the new method described in Section \ref{sec:7_pbo_nfs}, i.e.\ the price was calculated with $S_0=D_1$ and $S_0=D_2$ and we found the intersection between the line through these points and the straight line for the payoff $K-S_0$.

We calculated an approximate 95\% confidence interval (corresponding to 2 standard deviations) for the Monte-Carlo methods and Table \ref{tab:7_MC_Berm} shows that the results for our new methods described in Sections \ref{sec:7_pbo_grm} and \ref{sec:7_pbo_nfs}
are within this range for all cases tested. Furthermore, we can see that the results for Green's method and the new Spitzer based method are the same. More details about this Monte-Carlo scheme are included in Appendix \ref{sec:App_MC}.

\begin{table}[h] 
\centering
\begin{tabular}{rr|rr|rr|rr}
\hline
\hline
\multicolumn{8}{c}{Gaussian} \\
\hline
\multicolumn{2}{c}{Parameters}&\multicolumn{2}{|c}{Monte Carlo}&\multicolumn{2}{|c}{Spitzer}&\multicolumn{2}{|c}{Green}\\
\hline
$r$ & $\Delta t$ & price & 2 std dev & price & difference & price & difference\\
\hline
0.05 & 1 & 0.331801 & 8.41E-05&0.331811&-9.98E-06&0.331811&-9.98E-06\\
0.05 & 0.5 & 0.335196 & 8.92E-05 & 0.335223 &2.671E-05& 0.335223 &2.671E-05\\
0.05 &0.25 &0.336947 &9.00E-05 &0.336951 &-4.15E-06&0.336951 &-4.15E-06\\
0.05 &0.1 &0.337911 &1.66E-04 &0.337988&-7.746E-05&0.337988&-7.746E-05\\
\hline\hline
\multicolumn{8}{c}{VG} \\
\hline
\multicolumn{2}{c}{Parameters}&\multicolumn{2}{|c}{Monte Carlo}&\multicolumn{2}{|c}{Spitzer}&\multicolumn{2}{|c}{Green}\\
\hline
$r$ & $\Delta t$ & price & 2 std dev & price & difference& price & difference\\
\hline
0.05 &1 &0.120225 &2.15E-04 &0.120237 &-1.15E-05&0.120237 &-1.15E-05\\
0.05 &0.5 &0.123415&2.52E-04 &0.123298 &1.16E-04&0.123298 &1.16E-04\\
0.05 &0.25 &0.124996& 2.68E-04& 0.124919 &7.67E-05& 0.124919 &7.67E-05\\
0.05 &0.1 &0.125775 &2.54E-04& 0.125959 &-1.83E-04& 0.125959 &-1.83E-04\\
0.02 &1 &0.247040 &3.54E-04 &0.247078&-3.83E-05&0.247078&-3.83E-05\\
0.02 &0.1 &0.249747& 3.56E-04 &0.249756& -9.37E-06&0.249756& -9.37E-06\\
\hline\hline
\multicolumn{8}{c}{Merton jump-diffusion} \\
\hline
\multicolumn{2}{c}{Parameters}&\multicolumn{2}{|c}{Monte Carlo}&\multicolumn{2}{|c}{Spitzer}&\multicolumn{2}{|c}{Green}\\
\hline
$r$ & $\Delta t$ & price & 2 std dev & price & difference& price & difference\\
\hline
0.05 & 1& 0.119755 &2.58E-04 &0.119856 &-1.01E-04&0.119856 &-1.01E-04\\
0.05 &0.5 &0.123085 &2.39E-04 &0.122993 &9.21E-05&0.122993 &9.21E-05\\
0.05 &0.25& 0.124636 &2.77E-04& 0.124674& -3.82E-05& 0.124674& -3.82E-05\\
0.05 &0.1& 0.125925 &2.39E-04& 0.125767 &1.58E-04& 0.125767 &1.58E-04\\
\hline
\hline
\end{tabular}
\caption{Comparison between the value of a perpetual Bermudan option with $2^{20}$ price grid points and the value for the same contract using a Monte-Carlo approximation. Notice that the prices calculated using the new Spitzter based method and Green's method are the same and within two standard deviations of the Monte-Carlo price.}
\label{tab:7_MC_Berm}
\end{table}
\FloatBarrier
\section{Perpetual American options}\label{sec:PerAm}
As described by \cite{Green2010}, and implemented in \cite{Phelan2017Fluct}, we can exploit the relationship between Laplace and $z$-transforms described in Eq.~(\ref{eq:2_an_ztolap}) to extend the pricing methods from discrete to continuous monitoring (i.e.\ to perpetual American options). Unlike the previous examples for option pricing with continuous monitoring, where the application of option pricing was used as a motivating example for techniques which have relevance for other fields, the continuous (i.e.\ American) case is commonly used in financial contracts.
By defining $q=e^{-s\Delta t}$, we can write the continuously monitored equivalent to $\Phi(\xi,q)=1-q\Psi(\xi,\Delta t)$ as $\Phi_{\mathrm{c}}(\xi,s)=s-\psi(\xi)$, where $\psi(\xi)$ is the characteristic exponent of the characteristic function $\Psi(\xi,\Delta t)$.
%
%
Here, as $q=e^{-r\Delta t}$ then $s=r$. Both methods described in Sections \ref{sec:7_pbo_grm} and \ref{sec:7_pbo_nfs} are converted to continuous monitoring and we compare the results below.

\FloatBarrier
\subsection{Results for perpetual American options}
\begin{figure}[h]
\begin{center}
\includegraphics[width=0.48\textwidth]{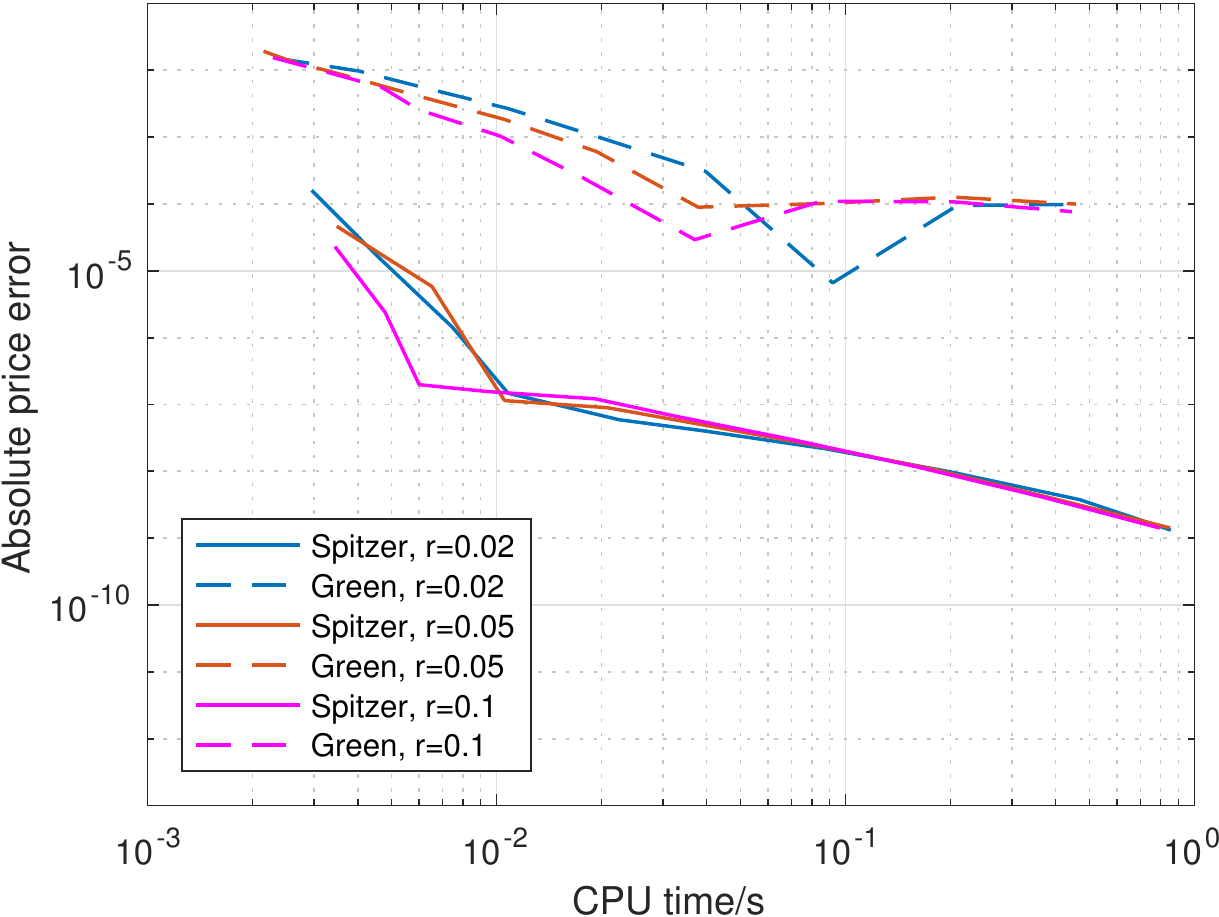}
\includegraphics[width=0.48\textwidth]{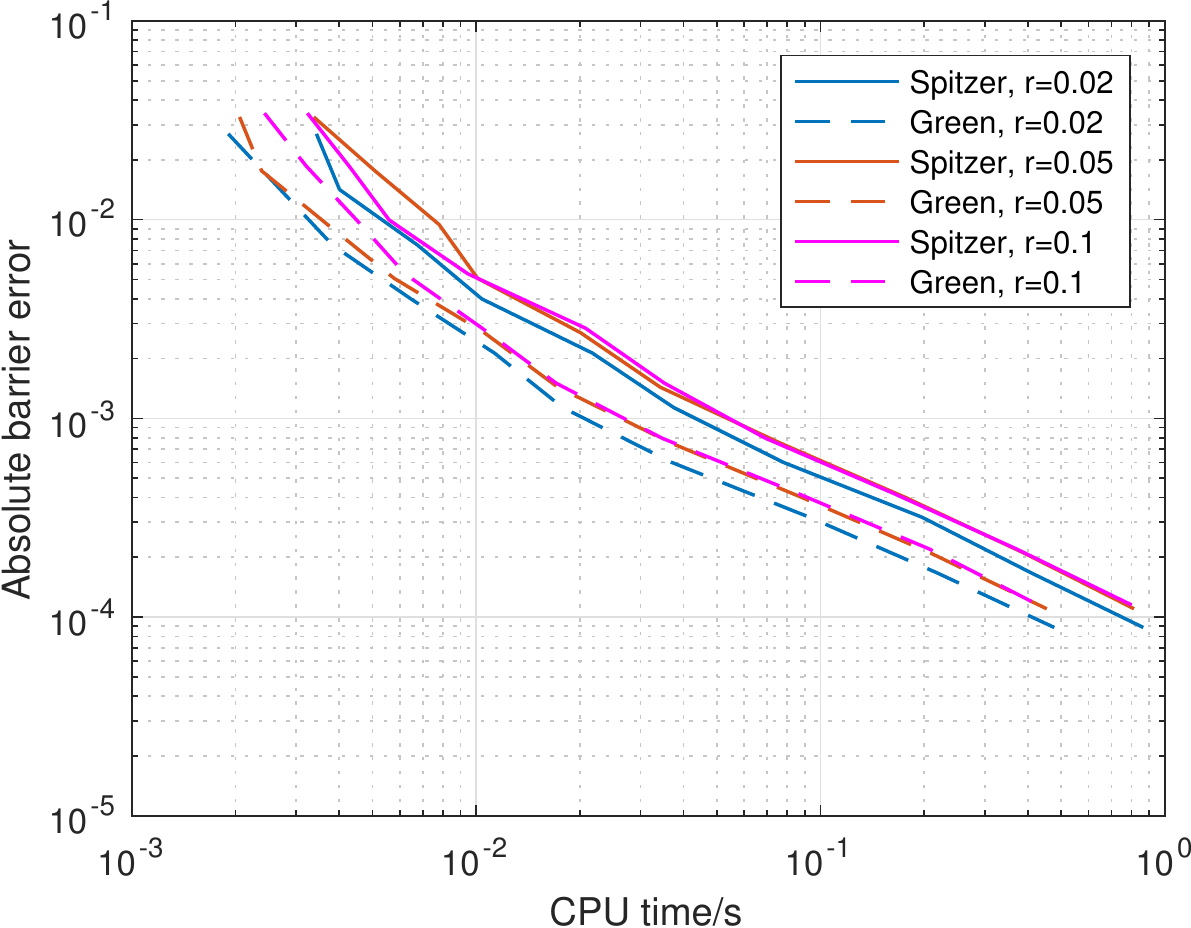}
\caption{Results for the price and barrier error convergence with CPU time with an underlying asset driven by a Gaussian process with a range of risk-free rates. The error is calculated using the closed-form expression by Merton as a reference. Notice that the price and barrier error convergence for the new method described in Section \ref{sec:7_pbo_nfs} labelled ``Spitzer" is faster than for the residue method described in Section \ref{sec:7_pbo_grm} labelled ``Green".}
\label{fig:7_pa_perpam_r_all_dt_allGaus}
\end{center}
\end{figure}

\begin{figure}[h]
\begin{center}
\includegraphics[width=0.48\textwidth]{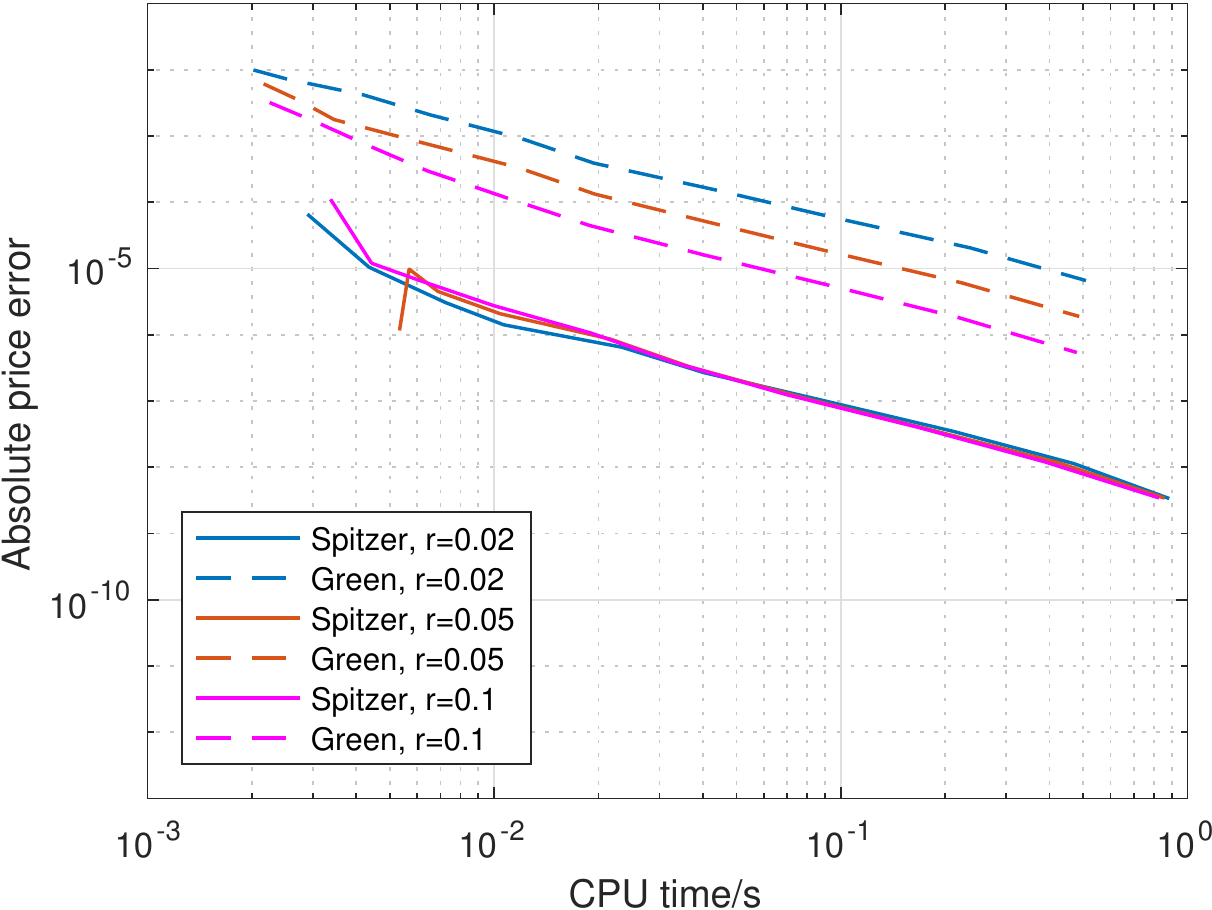}
\includegraphics[width=0.48\textwidth]{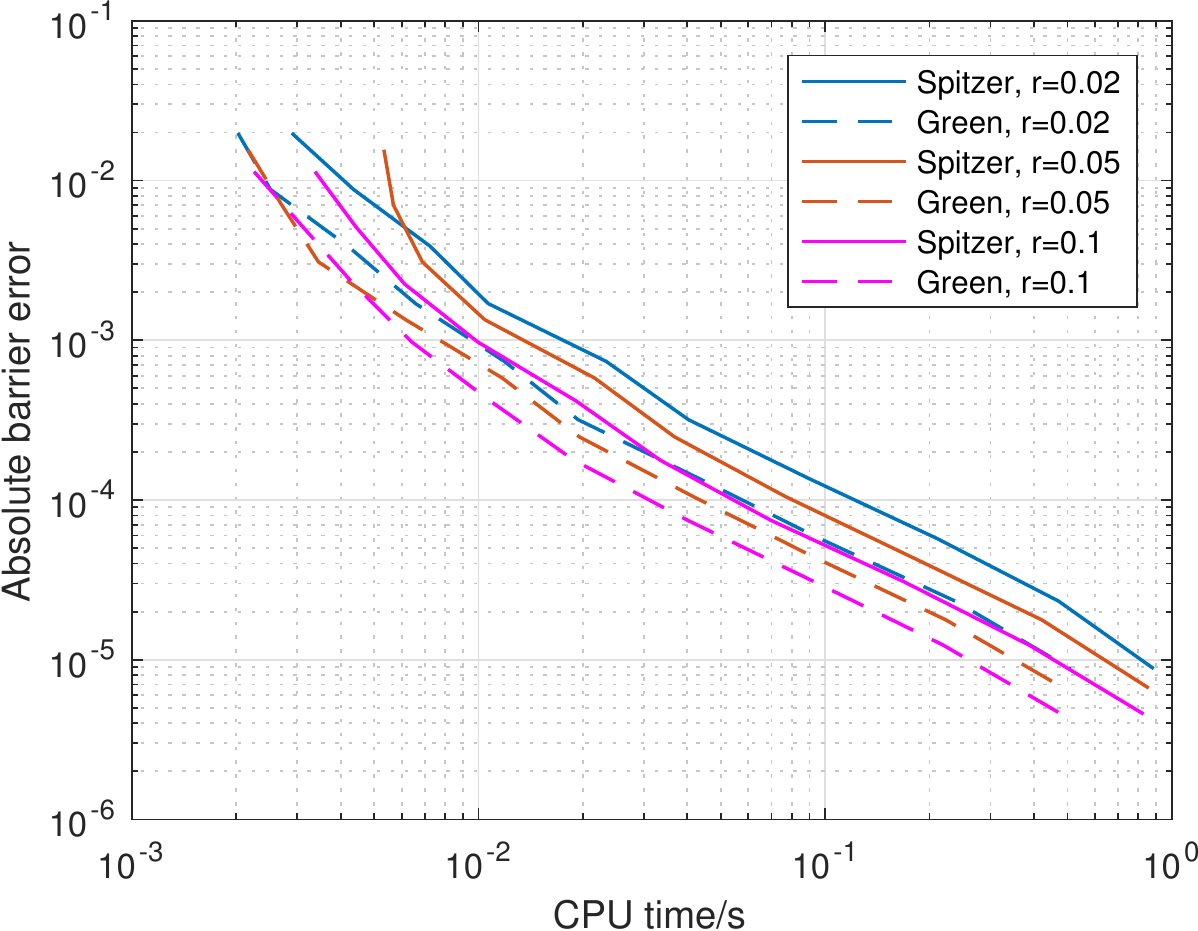}
\caption{Results for the price and barrier error convergence with CPU time with an underlying asset driven by a VG process with a range of risk-free rates. The error is calculated using the numerical result with the maximum grid size as a reference. Notice that the price and barrier error convergence for the new method described in Section \ref{sec:7_pbo_nfs} labelled ``Spitzer" is faster than for the residue method described in Section \ref{sec:7_pbo_grm} labelled ``Green".}
\label{fig:7_pa_perpam_r_all_dt_allVG}
\end{center}
\end{figure}

\begin{figure}
\begin{center}
\includegraphics[width=0.48\textwidth]{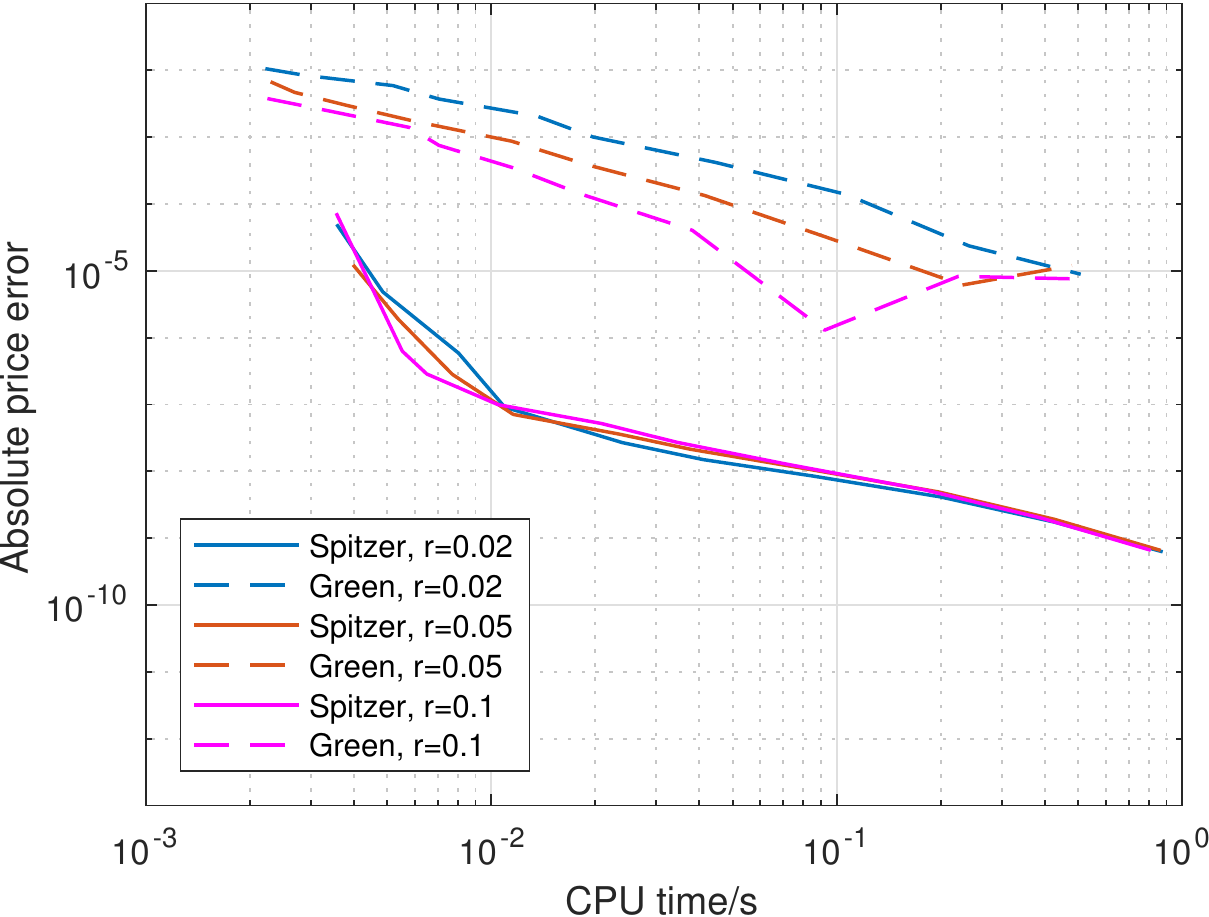}
\includegraphics[width=0.48\textwidth]{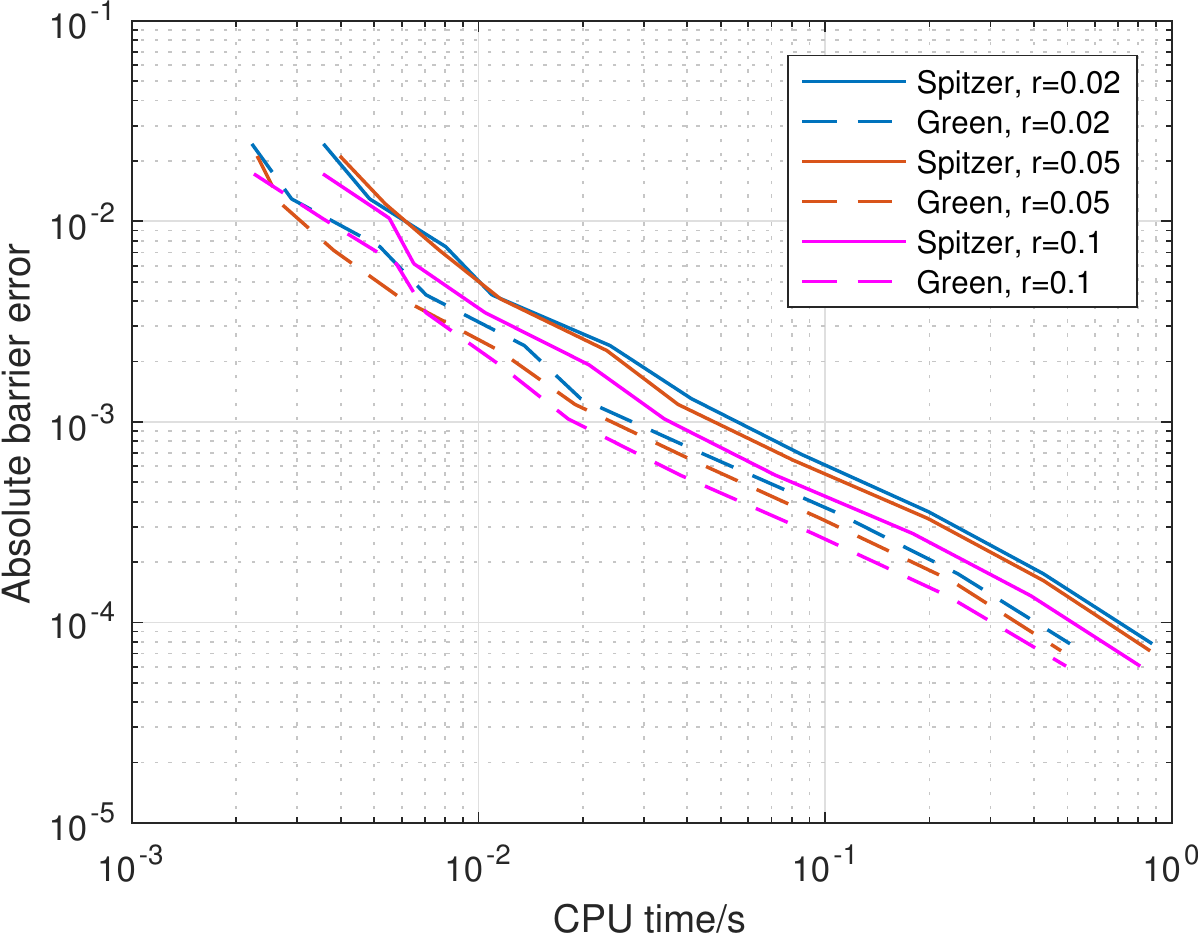}
\caption{Results for the price and barrier error convergence with CPU time with an underlying asset driven by a Merton process and a range of risk-free rates. The error is calculated using the numerical result with the maximum grid size as a reference. Notice that the price and barrier error convergence for the new method described in Section \ref{sec:7_pbo_nfs} labelled ``Spitzer" is faster than for the residue method described in Section \ref{sec:7_pbo_grm} labelled ``Green".}
\label{fig:7_pa_perpam_r_all_dt_allMerton}
\end{center}
\end{figure}

Figures \ref{fig:7_pa_perpam_r_all_dt_allGaus}--\ref{fig:7_pa_perpam_r_all_dt_allMerton} show results for both methods with the Gaussian, VG and Merton jump-diffusion processes. 

For assessing the accuracy of the methods with a Gaussian process we have the advantage over Bermudan options that closed-form formulas exist for the calculation of both the barrier and option price and so the results in Figure \ref{fig:7_pa_perpam_r_all_dt_allGaus} use the closed-form calculation from \cite{Merton1973} to calculate the error. For the other process we do not have a closed-form result, and the continuous nature of American options means that they cannot be accurately represented using Monte-Carlo methods which are inherently discrete. Therefore the absolute errors displayed in Figures \ref{fig:7_pa_perpam_r_all_dt_allVG} and \ref{fig:7_pa_perpam_r_all_dt_allMerton} are calculated against the result for the same method and process with the maximum number of FFT points.

In contrast to the results for perpetual Bermudan options, the performance of the two methods are very different. Our new Spitzer-based method is far superior, giving errors of approximately $10^{-6}$ in $10^{-2}$ seconds or less. For the Gaussian and Merton processes, Green's method fails to reach an error level of $10^{-6}$, and for the VG process it reaches this level about 100 times slower than our new method. 

\FloatBarrier
\subsection{Comparison between American and Bermudan option prices}
The performance for the direct calculation of the price of American options is sufficiently good for practical purposes, with the Spitzer method having an error of $~10^{-6}$ for a CPU time of $10^{-2}$s. 
However, it is of academic interest to study the use of the price for Bermudan options as an approximation to the price for American options. 

\begin{figure}[h]
\begin{center}
\includegraphics[width=0.48\textwidth]{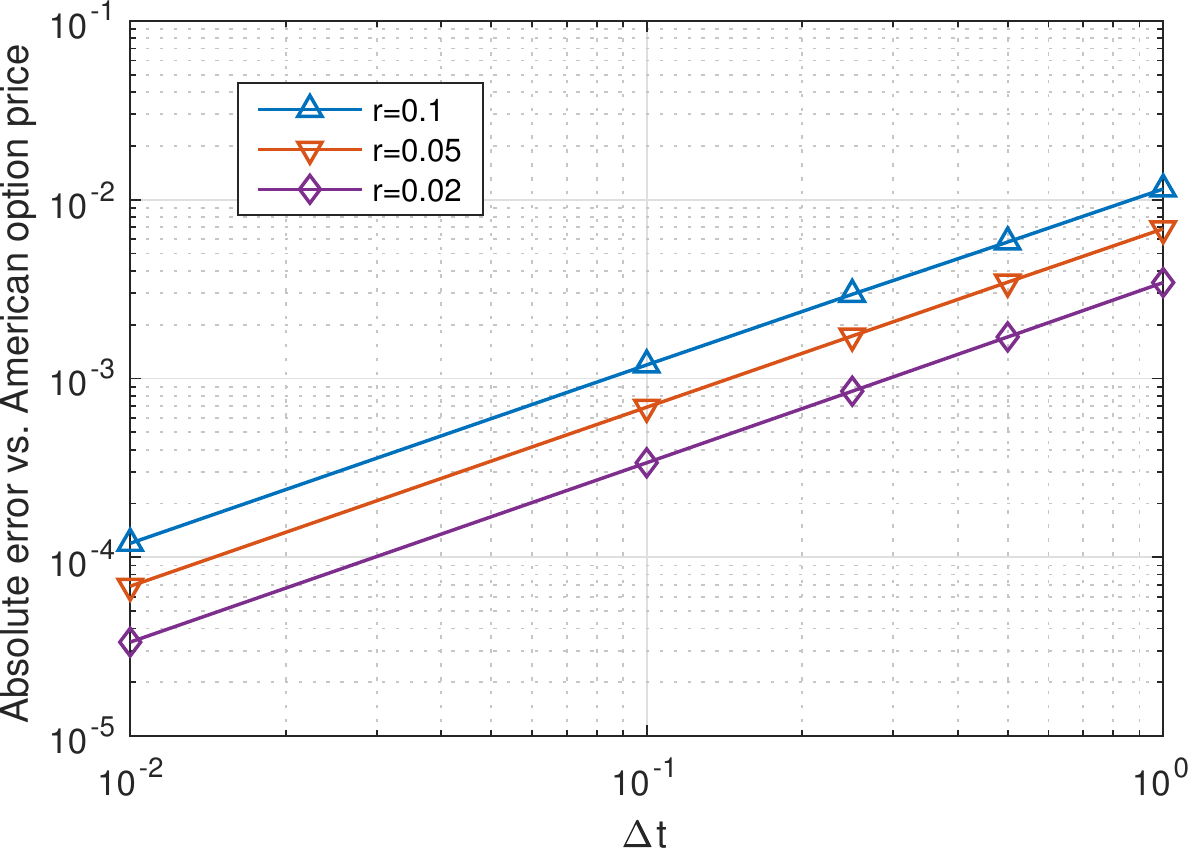}
\includegraphics[width=0.48\textwidth]{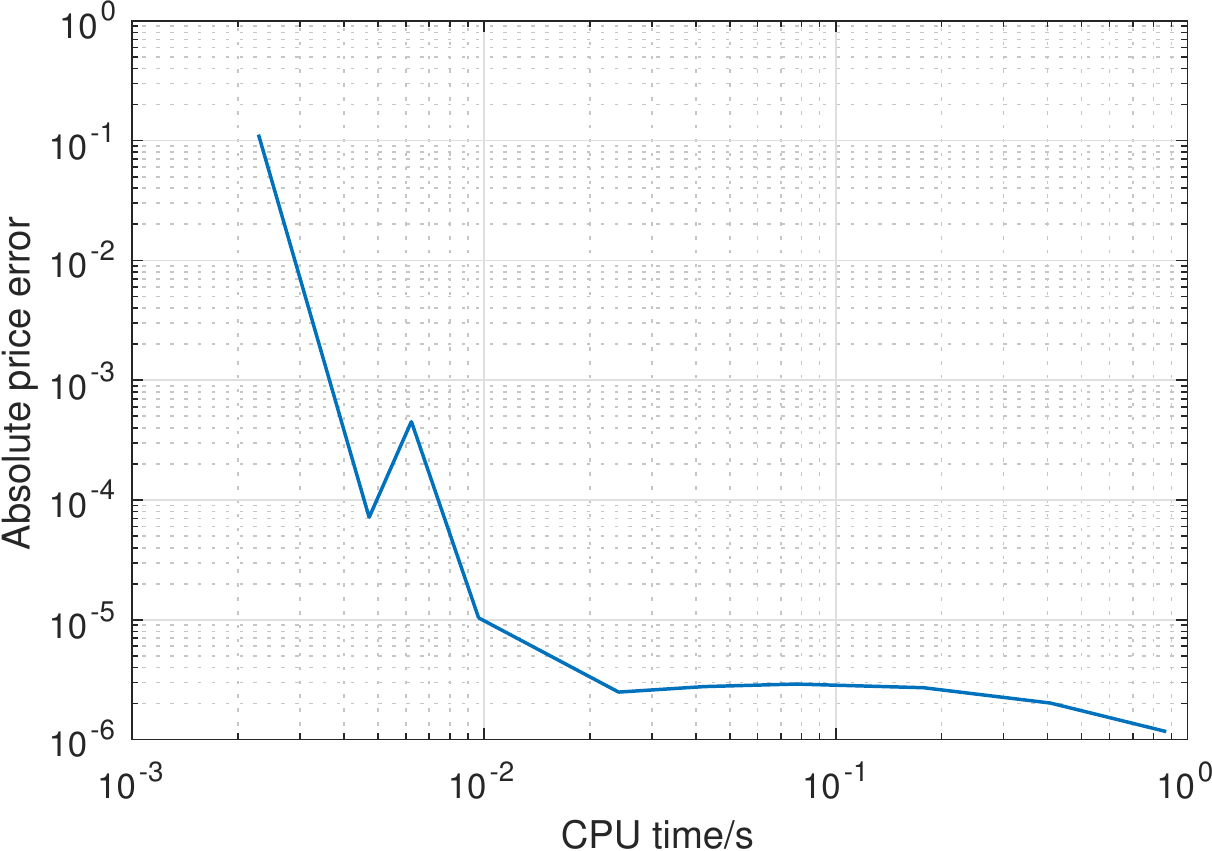}
\caption{Convergence of the price of perpetual Bermudan options to the price of perpetual American options. The left hand plot shows the convergence as $\Delta t\rightarrow 0$ with an underlying asset driven by a Gaussian process. The right hand side plot shows the error convergence of the price of a perpetual Bermudan option with $r=0.02$ and $\Delta t =$3E-06 vs.\ CPU time.}
\label{fig:7_pa_perpberm_r_0_02_dt_3Eminus6}
\end{center}
\end{figure}

The left hand plot in Figure \ref{fig:7_pa_perpberm_r_0_02_dt_3Eminus6} shows the price error compared to American options plotted against $\Delta t$ with an underlying asset driven by a Gaussian process and it is clear that the relationship is linear. By extrapolating this line, we can see that in order to achieve an error of $10^{-8}$ for $r=0.02$ a step size of $\Delta t=$0.3E-06 is required. The price convergence with this step size is shown in Figure \ref{fig:7_pa_perpberm_r_0_02_dt_3Eminus6} and we can see that reducing the step size this low destroys the monotonicity of the convergence and the excellent error performance that we were seeing for more realistic step sizes.

\FloatBarrier

\section{Conclusion}
We implemented three new pricing methods based on the Spitzer identities for pricing exotic options. 
We saw very fast error convergence with FFT grid size and, by extension, CPU time. Due to the low errors and the presence of the $10^{-11}$ error floor in the inverse $z$-transform, it was not possible to identify whether this error was exponentially or high order polynomially convergent. The extremely low errors and computational time make this a highly attractive method as it stands. However, it would be of academic interest to implement this method using an inverse $z$-transform with a lower error floor in order to determine the exact order of convergence which we achieve.

We presented a novel method for pricing perpetual Bermudan options which includes a new technique for directly computing the optimal exercise barrier and was based on the Spitzer identities and the sinc-based fast Hilbert transform. We also provide the first numerical implementation of the method designed by \cite{GreenThesis2009} which uses residue calculus to remove the requirement for the final inverse Fourier-$z$ transform.  Both methods performed well, but our new method showed significantly lower errors and CPU times, with a computational speed ten times faster for an error of $10^{-7}$. 

These methods were extended to perpetual American options (i.e.\ with continuous monitoring) and very different results were observed for the two methods. For Green's method, the factorisation error was very high and dominated the error convergence. However, for the new Spitzer based method, the factorisation error was much lower and therefore the effect only became dominant for errors below approximately $10^{-7}$. 

For errors greater than $10^{-7}$, the new Spitzer-based method for perpetual American options has errors at least as low as the method for perpetual Bermudan options so we conclude that, for practical purposes, there is no advantage in using discrete monitoring as an approximation for continuous monitoring. Comparing the two as an academic exercise showed that there is as a linear relationship between the size of the time step $\Delta t$ with discrete monitoring and the error compared to continuous monitoring. However, reducing $\Delta t$ to a sufficiently low value that the discrete method would be predicted to have a lower error than the continuous method significantly degraded the error convergence of the discretely monitored method such that there was no advantage gained.

\subsection*{Funding}

The support of the Economic and Social Research Council (ESRC) in funding the Systemic Risk Centre (grant number ES/K002309/1) and of the Engineering and Physical Sciences Research Council (EPSRC) in funding the UK Centre for Doctoral Training in Financial Computing and Analytics (grant number 1482817) are gratefully acknowledged.

\bibliographystyle{ormsv080}
\bibliography{quantilebermudan}
\newpage

\appendix
\noindent \textbf{\LARGE{Appendices}}
\section{Step-by-step numerical pricing procedures}\label{sec:App_proc}
Technical descriptions of the pricing schemes are described in Sections \ref{sec:Quantile}--\ref{sec:PerAm} of the main paper. Here we also provide detailed step-by-step procedures used for our MATLAB implementation.
\subsection{Numerical procedure for discretely monitored $\alpha$-quantile options} \label{sec:7_alphaq_Numproc}
In order to calculate the price of discretely monitored $\alpha$-quantile options using the Fourier-$z$ transform, we must express the time for the two independent random processes in terms of the number of monitoring dates. For $N$ monitoring dates we calculate $j=\alpha N$ to the nearest integer. The pricing procedure is then
\begin{enumerate}
\item Compute the characteristic function $\Psi(\xi+i\alpha_{\mathrm{d}},\Delta t)$ of the underlying transition density, where $\alpha_{\mathrm{d}}$ is the damping parameter introduced in Section \ref{sec:fourhilb}, Eq.~(\ref{eq:Planch}) and $\Delta t = T/N$.\label{step:7_alphaq1}
\item Use the Plemelj-Sokhotsky relations with the sinc-based Hilbert transform to factorise
\begin{equation}
\label{eq:Phifactq}
\Phi(\xi,q):=1-q\sigma(\xi/\xi_{\max})\Psi(\xi+i\alpha_\mathrm{d},\Delta t)=\Phi_{\oplus}(\xi,q)\Phi_{\ominus}(\xi,q),
\end{equation}
where $\sigma(\eta)$ is an exponential filter of order 12 as defined by \cite{gottlieb1997gibbs} and $q$ is selected according to the criteria specified by \cite{Abate1992} for the inverse $z$-transform.
\item Calculate \label{step:7_alphaq2}
\begin{align}
\widetilde{\widehat{p}}_{X_M}(\xi,q)&=\frac{1}{\Phi_{\oplus}(\xi,q)\Phi_{\ominus}(0,q)},\\
\widetilde{\widehat{p}}_{X_m}(\xi,q)&=\frac{1}{\Phi_{\oplus}(0,q)\Phi_{\ominus}(\xi,q)}.
\end{align}
\item Apply the inverse $z$-transform for $j$ and $N-j$ dates respectively \label{step:7_alphaq4}
\begin{align}
\widehat{p}_{X_M}(\xi,j)&=\mathcal{Z}^{-1}_{q\rightarrow j}\left[\widetilde{\widehat{p}}_{X_M}(\xi,q)\right],\\
\widehat{p}_{X_m}(\xi,N-j)&=\mathcal{Z}^{-1}_{q\rightarrow N- j}\left[\widetilde{\widehat{p}}_{X_m}(\xi,q)\right].
\end{align}
\item Calculate the Fourier transform of the required probability distributions for $X_M$ and $X_m$ by obtaining the real parts of $\widehat{p}_{X_M}(\xi,j)$ and $\widehat{p}_{X_m}(\xi,N-j)$ directly in the Fourier domain using
\begin{align}
\widehat{p}^\mathrm{\,Re}_{X_M}(\xi,j):=\mathcal{F}_{x\to\xi}\big[\mathrm{Re}\,p_{X_M}(x,j)\big]&=\frac{1}{2}\big[\widehat{p}_{X_M}(\xi,j)+\widehat{p}_{X_M}^{\,*}(-\xi,j)\big],\\
\widehat{p}^\mathrm{\,Re}_{X_m}(\xi,N-j):=\mathcal{F}_{x\to\xi}\big[\mathrm{Re}\,p_{X_m}(x,j)\big]&=\frac{1}{2}\big[\widehat{p}_{X_m}(\xi,N-j)+\widehat{p}_{X_m}^{\,*}(-\xi,N-j)\big].
\end{align}
where $\cdot^*$ denotes the complex conjugate.
\item Calculate the Fourier transform of the probability distribution for $X_\alpha$ over N monitoring dates as
\begin{align}
\widehat{p}_{X_\alpha}(\xi,N)=\widehat{p}^\mathrm{\,Re}_{X_M}(\xi,j)\widehat{p}^\mathrm{\,Re}_{X_m}(\xi,N-j).
\end{align}
\item Calculate the price of the discretely monitored $\alpha$-quantile option as
\begin{align}
v(0,0)=\mathcal{F}^{-1}_{\xi\rightarrow x}\big[\sigma(\xi/\xi_{\max})\widehat{p}_{X_\alpha}(\xi,N)\widehat{\phi}^*(\xi)\big](0),
\end{align}
where $\sigma(\eta)$ is an exponential filter of order 12, as before.
\end{enumerate}

\subsection{Numerical procedure for Green's residue method for perpetual Bermudan options}\label{sec:7_Berm_Green}
\begin{enumerate}
\item Compute the characteristic function $\Psi(\xi+i\alpha_{\mathrm{d}})$, where $\alpha_{\mathrm{d}}$ is the damping parameter.\label{step:7_Berm1}
\item Use the Plemelj-Sokhotsky relations with the sinc-based Hilbert transform to factorise\label{step:7_Berm2}
\begin{equation}
\Phi(\xi,q):=1-e^{-r\Delta t}\sigma(\xi/\xi_{\max})\Psi(\xi+i\alpha_{\mathrm{d}},\Delta t)=\Phi_{\oplus}(\xi,e^{-r\Delta t})\Phi_{\ominus}(\xi,e^{-r\Delta t}),
\end{equation}
where $\sigma(\eta)$ is an exponential filter of order 12, as defined in \cite{gottlieb1997gibbs}. From this we can directly obtain $\Phi_{\ominus}(0,e^{-r\Delta t})$.
\item Use the shift theorem to calculate $\Phi_{\ominus}(i,e^{-r\Delta t})$ as
\begin{equation}
\Phi_{\ominus}(i,e^{-r\Delta t}):=\mathcal{F}_{x\rightarrow \xi}\left[e^{-x}\mathcal{F}^{-1}_{\xi\rightarrow x}\left[\Phi_{\ominus}(\xi,e^{-r\Delta t})\right]\right](0).
\end{equation}
\item Calculate the optimal exercise boundary
\begin{align}
D_{\mathrm{opt}}:=K\frac{\Phi_{\ominus}(0,e^{-r\Delta t})}{\Phi_{\ominus}(i,e^{-r\Delta t})}
\end{align}
and compute $l_{\mathrm{opt}}=\log(D_{\mathrm{opt}}/S_0)$.
\item Decompose $P(\xi,e^{-r\Delta t})$ around $l_{\mathrm{opt}}$
\begin{align}
P(\xi,e^{-r\Delta t})=P_{l_{\mathrm{opt}}+}(\xi,e^{-r\Delta t})+P_{l_{\mathrm{opt}}-}(\xi,e^{-r\Delta t})
\end{align}
and directly obtain $P_{l_{\mathrm{opt}}-}(0,e^{-r\Delta t})$.
\item Use the shift theorem to calculate $P_{l_{\mathrm{opt}}-}(i,e^{-r\Delta t})$ as
\begin{equation}
P_{l_{\mathrm{opt}}-}(i,e^{-r\Delta t}):=\mathcal{F}_{\xi\rightarrow \xi}\left[e^{-x}\mathcal{F}^{-1}_{\xi\rightarrow x}\left[P_{l_{\mathrm{opt}}-}(0,e^{-r\Delta t})\right]\right](0).
\end{equation}
\item Calculate the option price as 
\begin{equation}
v(0,0)=K \Phi_{\ominus}(0,e^{-r\Delta t})\left[P_{l_{\mathrm{opt}}-}(0,e^{-r\Delta t})-P_{l_{\mathrm{opt}}-}(i,e^{-r\Delta t})\right].
\end{equation}
\end{enumerate}

\subsection{Numerical procedure for new Spitzer based method for perpetual Bermudan options}\label{sec:7_Berm_Spitz}
\begin{enumerate}
\item Compute the characteristic function $\Psi(\xi+i\alpha_{\mathrm{d}})$, where $\alpha_{\mathrm{d}}$ is the damping parameter.
\item Use the Plemelj-Sokhotsky relations with the sinc-based Hilbert transform to factorise
\begin{equation}
\Phi(\xi,q):=1-e^{-r\Delta t}\sigma(\xi/\xi_{\max})\Psi(\xi+i\alpha_{\mathrm{d}},\Delta t)=\Phi_{\oplus}(\xi,e^{-r\Delta t})\Phi_{\ominus}(\xi,e^{-r\Delta t}),
\end{equation}
where $\sigma(\eta)$ is an exponential filter of order 12. 
\item Decompose $P(\xi,e^{-r\Delta t})$ around the grid step $l_\epsilon$ in the log-price domain,
\begin{align}
P(\xi,e^{-r\Delta t})=P_{l_\epsilon+}(\xi,e^{-r\Delta t})+P_{l_\epsilon-}(\xi,e^{-r\Delta t}).
\end{align}
\item Calculate the PDF for the calibration
\begin{align}
\widehat{p}_{l_\epsilon-}(\xi,e^{-r\Delta t})=P_{l_\epsilon-}(\xi,e^{-r\Delta t})\Phi_{\ominus}(\xi,e^{-r\Delta t}).
\end{align}
\item Setting $D_1$ = 0 and $D_2\in(0,K)$, calculate
\begin{gather}
v_{\mathrm{D_1}}(0,0) = K\widehat{p}_{l_\epsilon-}(0,e^{-r\Delta t})\\
v_{\mathrm{D_2}}(0,0) = \mathcal{F}^{-1}_{\xi\rightarrow x}\big[\widehat{\phi}^*(\xi)\widehat{p}_{l_\epsilon-}(0,e^{-r\Delta t})\big](0),
\end{gather}
where $\widehat{\phi}(\xi)$ is the Fourier transform of the damped payoff $\phi(x)$ with $x=\log(S(t)/D_2)$.
\item Calculate the optimal exercise barrier
\begin{equation}
D_{\mathrm{opt}} = \frac{\left[K-v_{\mathrm{D_1}}(0,0)\right]D_2}{v_{\mathrm{D_2}}(0,0)-v_{\mathrm{D_1}}(0,0)+D_2}
\end{equation}
and compute $l_{\mathrm{opt}}=\log(D_{\mathrm{opt}}/S_0)$.
\item Decompose $P(\xi,e^{-r\Delta t})$ around $l_{\mathrm{opt}}$
\begin{align}
P(\xi,e^{-r\Delta t}) = P_{l_{\mathrm{opt}}+}(\xi,e^{-r\Delta t})+P_{l_{\mathrm{opt}}-}(\xi,e^{-r\Delta t}).
\end{align}
\item Calculate the PDF for the price
\begin{align}
\widehat{p}_{l_{\mathrm{opt}}-}(\xi,e^{-r\Delta t})=P_{l_{\mathrm{opt}}-}(\xi,e^{-r\Delta t})\Phi_{\ominus}(\xi,e^{-r\Delta t}).
\end{align}
\item Calculate the option price
\begin{align}
v(0,0)=\mathcal{F}^{-1}_{\xi\rightarrow x}\big[\widehat{\phi}^*(\xi)\widehat{p}_{l_{\mathrm{opt}}}(0,e^{-r\Delta t})\big](0),
\end{align}
where $\widehat{\phi}(\xi)$ is the Fourier transform of the damped payoff $\phi(x)$ with $x=\log(S(t)/S_0)$.
\end{enumerate}

\subsection{Pricing procedure for perpetual American options}
For both methods we adapt the pricing procedure for discretely monitored options to continuous monitoring by replacing Steps \ref{step:7_Berm1}--\ref{step:7_Berm2} in the procedures described in Sections \ref{sec:7_Berm_Green} and \ref{sec:7_Berm_Spitz} with
\begin{enumerate}
\item Compute the characteristic exponent $\psi(\xi+i\alpha_{\mathrm{d}})$ of the underlying transition density. \label{step:7_amcont1}
\item Use the Plemelj-Sokhotsky relations with the sinc-based Hilbert transform to factorise
\begin{equation}
\label{eq:Phifacta}
\Phi_{\mathrm{c}}(\xi,r):=r-\psi(\xi+i\alpha_{\mathrm{d}})=\Phi_{\mathrm{c}\oplus}(\xi,r)\Phi_{\mathrm{c}\ominus}(\xi,r)
\end{equation}
where $r$ is the risk-free rate.
\end{enumerate}
We then continue with the calculations in Section \ref{sec:7_Berm_Green} or \ref{sec:7_Berm_Spitz} as before but with $\Phi_{\mathrm{c}}(\xi,r)$ in place of $\Phi(\xi,e^{-r\Delta t})$.

\section{Process parameters}\label{sec:App_params}
Table \ref{tab:A_Parasetupalpha_earlyex} contains the process parameters used for the numerical tests.
\begin{table}[ht]
\begin{center}
\begin{tabular}{lllr}
\hline\hline
Model & $\Psi(\xi,t)$ & Symbol & Value\\
\hline
\multirow{5}{*}{Kou}&\multirow{5}{*}{$e^{-t\left(\frac{\sigma^2\xi^2}{2}-\lambda\left(\frac{(1-p)\eta_2}{\eta_2+i\xi}+\frac{p\eta_1}{\eta_1-i\xi}-1\right)\right)}$} & $p$ & 0.3\\
& & $\lambda$ & 3\\
& & $\sigma$ & 0.1\\
& &$\eta_1$ & 40\\
& & $\eta_2$ & 12\\
\hline
\multirow{3}{*}{VG}&\multirow{3}{*}{$(1-i\nu\xi\theta+\nu\sigma^2\xi^2/2)^{-t/\nu}$} & $\theta$ & $\frac{1}{9}$\\
\noalign{\vskip 0.5mm}
& & $\sigma$ &$\frac{1}{3\sqrt{3}}$\\
\noalign{\vskip 0.5mm}
& & $\nu$ & 0.25 \\
\hline
\multirow{4}{*}{MJD}&\multirow{4}{*}{$e^{-\frac{1}{2}\sigma^2\xi^2+\lambda(e^{i\alpha_m\xi-\frac{1}{2}\delta^2\xi^2}-1)}$} & $\sigma$ & 0.1\\
\noalign{\vskip 0.5mm}
& & $\lambda$ &3\\
\noalign{\vskip 0.5mm}
& & $\alpha_m$ & -0.05 \\
\noalign{\vskip 0.5mm}
& & $\delta$ & 0.086\\
\hline\hline
\end{tabular}
\end{center}
\caption{Parameters for the numerical tests; $\Psi(\xi,t)$ is the characteristic function of the process that models the log return of the underlying asset.}
\label{tab:A_Parasetupalpha_earlyex}
\end{table}

\section{Monte-Carlo pricing procedures}\label{sec:App_MC}
We describe in more detail the Monte-Carlo pricing procedures which were used to verify the prices calculated by the numerical pricing procedures in this paper.
\subsection{$\alpha$-quantile options}\label{sec:MC_alpha_quant}
For underlying assets modelled by Gaussian, VG and Merton jump-diffusion processes, these have paths which can be simply constructed using the built-in Matlab commands for normal, gamma and Poisson random variables. In all cases the risk neutral drift was calculated using $\mu_{\mathrm{RN}}=\psi(-i)$ as described by \cite{Feng2008}, where $\psi(\xi)$ is the characteristic exponent of the L\'evy process. 

\paragraph{VG process.} A single step over time $\Delta t$ was simulated using the following procedure:
\begin{enumerate}
\item Calculate the gamma-distributed random variable $\Delta G$ which has a probability distribution $\Gamma(\frac{1}{\nu \Delta t},\nu)$.
\item Calculate the process step
\begin{equation}
\Delta X=\mu_{\mathrm{RN}}\Delta t+\theta \Delta G+\sigma\sqrt{\Delta G}\zeta,
\end{equation}
where $\zeta$ is a standard normal random variable.
\end{enumerate}

\paragraph{MJD process.} A single step over time $\Delta t$ was simulated using the following procedure:
\begin{enumerate}
\item Calculate the Poisson distributed random variable with parameter $\lambda\Delta t$.
\item Calculate the process step:
\begin{equation}
\Delta X=\mu_{\mathrm{RN}}\Delta t+\sigma\sqrt{\Delta t}\zeta_1+\mu_{\mathrm{J}}N+\sigma_{\mathrm{J}}\sqrt{N}\zeta_2,
\end{equation}
where $\zeta_1$ and $\zeta_2$ are independent standard normal random variables.
\end{enumerate}

We used two methods to calculate the value for the $\alpha$-quantile. For the first we simulated a path of $N$ points and then found the $j^{\mathrm{th}}$ smallest value. For the second we combined the Dassios-Port-Wendel identity with the Monte-Carlo method as in \cite{Ballotta2001}. Two independent paths of length $\alpha N$ and $(1-\alpha)N$ dates are simulated and the sum of their respective infimum and supremum are calculated to provide an estimate of the $\alpha$-quantile.

\subsection{Perpetual Bermudan options}
We wrote Monte-Carlo pricing procedures with underlying assets driven by Gaussian, VG and Merton jump-diffusion processes using the same techniques to simulate the paths as described in Section \ref{sec:MC_alpha_quant}.

Although the discrete nature of perpetual Bermudan options is appropriate for a Monte-Carlo simulation, the absence of an expiry date means that a Monte-Carlo scheme with a finite number of dates will not represent the contract exactly. However, we truncate the Monte-Carlo simulation so far in the future that the effect of disregarding these dates is less than the standard deviation of the Monte-Carlo method itself. (Clearly this is more feasible for high discount factors and large time steps as the effect of future dates is discounted away more rapidly.) Finding the optimal exercise barrier uses the same philosophy as the new method described in Section \ref{sec:7_pbo_nfs}, i.e.\ the price was calculated with $S_0=D_1$ and $S_0=D_2$ and we found the intersection between the line through these points and the straight line for the payoff $K-S_0$.

\end{document}